\documentclass[useAMS,usenatbib,letterpaper]{mn2e}
\usepackage{amssymb}
\usepackage{amsmath}
\usepackage{subfigure}
\usepackage{float}
\usepackage{appendix}
\usepackage{multirow}
\usepackage{graphicx}

\title{Alignment of Brightest Cluster Galaxies with their Host Clusters} 
\author[M.Niederste-Ostholt et al.]{Martin Niederste-Ostholt$^{1}$, Michael A. Strauss$^{2}$, Feng Dong$^{2}$,
\newauthor Benjamin P. Koester$^{3}$, Timothy A. McKay$^{4}$\\
$^{1}$Institute of Astronomy, Madingley Rd, Cambridge, CB3 0HA, UK; mno@ast.cam.ac.uk\\
$^{2}$Department of Astrophysical Sciences, Princeton University, Princeton, NJ 08544, USA; strauss, feng@astro.princeton.edu\\
$^{3}$Department of Astronomy and Astrophysics, University of Chicago, Chicago, IL 60637, USA; bkoester@oddjob.uchicago.edu\\
$^{4}$Department of Physics, University of Michigan, Ann Arbor, MI 48109, USA; tamckay@umich.edu}

\begin{document}

\date{February 2010}

\pagerange{\pageref{firstpage}--\pageref{lastpage}} \pubyear{2010}

\maketitle

\label{firstpage}

\begin{abstract}
We examine the alignment between Brightest Cluster Galaxies (BCGs) and their host clusters in a sample of $7031$ clusters with $0.08<z<0.44$ found using a matched-filter algorithm and an independent sample of $5744$ clusters with $0.1<z<0.3$ selected with the maxBCG algorithm, both extracted from the Sloan Digital Sky Survey Data Release 6 imaging data. We confirm that BCGs are preferentially aligned with the cluster's major axis; clusters with dominant BCGs ($>0.65$ mag brighter than the mean of the second and third ranked galaxies) show stronger alignment than do clusters with less dominant BCGs at the $4.4\sigma$ level. Rich clusters show a stronger alignment than do poor clusters at the $2.3\sigma$ level. Low redshift clusters ($z<0.26$) show more alignment than do high redshift ($z>0.26$) clusters, with a difference significant at the $3.0\sigma$ level. Our results do not depend on the algorithm used to select the cluster sample, suggesting that they are not biased by systematics of either algorithm. The correlation between BCG dominance and cluster alignment may be a consequence of the hierarchical merging process which forms the cluster. The observed redshift evolution may follow from secondary infall at late redshifts.
\end{abstract}

\section{Introduction}
\indent The present cosmological paradigm \citep{Komatsu:2009p1215} predicts the existence of coherent structures in the universe as well as alignments between structures on various scales. In a cold dark-matter dominated universe with structures forming hierarchically from the bottom up, filamentary structure formation and tidal forces may cause alignments between different elements of the hierarchy \citep[e.g.][]{Torlina:2007p58,Faltenbacher:2008p67}. Simulations using a $\Lambda$CDM cosmology have predicted alignments between clusters and super-clusters \citep[e.g.][]{Basilakos:2006p69}, alignments between clusters \citep{Hopkins:2005p348}, and alignments between galaxies and clusters \citep{Dubinski:1998p366}. Observational studies have found alignments between galaxies and large-scale structures \citep{Faltenbacher2009RAA.....9...41F}; clusters and clusters \citep[e.g..][]{Binggeli:1982p376}, galaxies and clusters \citep[e.g..][]{Sastry:1968p242,Binggeli:1982p376,Struble:1990p48,Kim:2002ASPC..268..395K,Hashimoto:2008MNRAS.390.1562H}, and galaxies within groups \citep[e.g..][]{Yang:2006p82,Faltenbacher:2007p60,Wang:2008p65}. Most of the earlier observational probes of galaxy-cluster alignment were based on small samples and only have shallow redshift coverage. With the availability of large imaging surveys such as the Sloan Digital Sky Survey \citep[SDSS][]{York:2000p367} we can now extend these earlier observations. In this spirit we study the alignment of brightest cluster galaxies (BCGs) with their host clusters from SDSS data in this paper.
\\
\indent BCGs are not simply the brightest galaxy picked from the luminosity function of a cluster. \citet{Humason:1956p416,Tremaine:1977p403,Lauer:1988p402,Postman:1995p237,Garijo:1997p54,Dubinski:1998p366,Loh:2006p425} and others point out that BCGs have properties that suggest a formation scenario distinct from that of other galaxies, including photometric and colour homogeneity, a metric luminosity independent of cluster richness and significantly higher than that of lower-ranked galaxies, disturbed morphologies, extended low surface brightness stellar haloes and central location.
\\
\indent The shapes of clusters, the special properties of brightest cluster galaxies, and their alignment are clues to the process by which clusters formed. Understanding the alignment offers an observational test of the current cosmological paradigm. The alignment of BCGs with their hosts was first noted by \citet{Sastry:1968p242} and studied by \citet{Carter:1980p229}. Detailed observations were carried out by \citet{Binggeli:1982p376} (hence the alignment is also known as the Binggeli effect); \citet{Struble:1985p234} (who found no alignment with a small cluster sample); \citet{Rhee:1987p231,Djorgovski1987,Lambas:1988p43,Struble:1990p48,Trevese:1992p39,Fuller:1999p61,Kim:2002ASPC..268..395K,Donoso:2006p66} and \citet{Siverd:2009p516}.  The effect is observed when the cluster shape is defined by the distribution of member galaxies or by the observed distribution of x-ray emitting hot gas  \citep{Hashimoto:2008MNRAS.390.1562H}. The later studies were able to confirm the alignment for clusters with redshifts up to $z=0.5$. The effect is strongest for cluster shapes measured for red, centrally concentrated galaxies, and it weakens with increasing separation between the central galaxy and the satellites.  
\\
\indent The two leading candidate mechanisms to explain the alignment of the BCG with their parent cluster are that both are formed from infall along preferred directions in filaments \citep[e.g.][]{Dubinski:1998p366}, and that BCGs are aligned by tidal interactions \citep[e.g.][]{Faltenbacher:2008p67}. The points in cosmic history at which these mechanisms act are very different: whereas filamentary infall is likely to affect alignments during and immediately after cluster virialization tidal affects can act during the cluster's entire lifetime. \citet{Ciotti:1994p478} have shown that the time scale on which a prolate galaxy's orientation is affected by a cluster's tidal field is much shorter than a Hubble time. Present-day alignments may therefore either be the result of primordial alignments stemming from the period of cluster formation or can have grown during the cluster's lifetime. By studying the redshift evolution of the alignment effect we can hope to distinguish these two cases. 
\\
\indent In this paper we use Sloan Digital Sky Survey Data Release 6 data \citep{York:2000p367,AdelmanMcCarthy:2008p362} to study the alignment effect in $12,755$ clusters extending out to $z=0.44$. In \S 2 we describe our cluster selection and in \S 3 we analyse the dependence of alignment on BCG dominance, cluster richness, and redshift. We discuss the implications of our findings in \S 4 and summarise in \S 5. Throughout this paper we assume an $\Omega_m=0.3$, $\Omega_{\Lambda}=0.7$, and $H_0 = 70$ km s$^{-1}$ Mpc$^{-1}$ cosmology.

\section{Data}
\subsection{The Sloan Digital Sky Survey}
\indent The Sloan Digital Sky Survey \citep[SDSS;][]{York:2000p367} is an imaging and spectroscopic survey that has covered one-quarter of the celestial sphere. It is carried out with a dedicated 2.5m telescope \citep{2006AJ....131.2332G} using a large format CCD camera \citep{Gunn:1998p1435}.  The imaging data are collected in drift-scan mode in five band passes ($u$,$g$,$r$,$i$,$z$) with effective wavelengths of $3551,\;4686,\;6165,\;7481$, and $8931$ \AA. The imaging data are automatically reduced through a series of software pipelines which find and measure objects and provide photometric and astrometric calibrations \citep{Lupton:2001p1436,Lupton:2002SPIE.4836..350L,Pier:2003p1437,Tucker:2006p1438}. The photometric calibrations are accurate to about $1\%$ rms in $g$, $r$, and $i$, $3\%$ in $u$ and $2\%$ in $z$ for bright ($< 20$ mag) point sources \citep{Ivezic:2004p1443}. We restrict ourselves to the objects that have reliable photometric data by using the SDSS clean photometry flags. Targets for spectroscopy are selected from the imaging data on the basis of their photometric properties. A pair of dual fibre-fed spectrographs \citep{Uomoto1999} can observe 640 spectra at a time with a wavelength coverage of $3800$ \AA$\;$to $9200$ \AA$\;$and a resolution of $1800$ \AA$\;$to $2100$ \AA. 
\\
\indent The SDSS data are described in the data release papers \citep[][for the Sixth Data Release which we use]{AdelmanMcCarthy:2008p362} and are documented at \textbf{http://www.sdss.org}.

\subsection{Cluster Catalogues}
\indent In this paper we use two catalogues of clusters of galaxies. One by \citet{Koester:2007p484,Koester:2007p485} takes advantage of the concentration of cluster galaxies in colour-magnitude space on the red sequence, and the other, by \citet{Dong:2008p220} identifies clusters by matching galaxy distributions in position-magnitude space to an a priori filter. The two catalogues are likely to have different systematic errors such as miscentering and problems caused by cluster overlap, so computing results from the two allows us to reduce the severity such problems are.
\\
\indent There is a well known bimodality of galaxies in colour-colour space \citep[e.g.][]{Strateva:2001p504,Baldry:2004p481,Bell:2004p497,Cassata:2008p507} with red quiescent spheroidal galaxies occupying what is known as the red sequence, and blue actively star-forming galaxies concentrated in the blue cloud.  As first noted by \citet{Baum:1959PASP...71..106B}, red sequence galaxies are predominantly found in clusters of galaxies. Since the location of the red sequence in colour-colour space is dependent on redshift via the K-correction it is a viable photometric redshift indicator for galaxy clusters and field galaxies \citep[e.g..][and references therein]{Annis:1999AAS...195.1202A,Gladders:2000p360,Koester:2007p484,Koester:2007p485,Eisenhardt:2008p1444}. These properties make the red sequence an ideal tool for identifying galaxy clusters in surveys \citep[e.g..][]{Yee:1999ASPC..191..166Y,Gladders:2000p360,Lubin:2000ApJ...531L...5L,Gladders:2005ApJS..157....1G,Koester:2007p484}. The red sequence allows one to isolate galaxies at a given redshift and thus remove contamination by projected foreground and background galaxies.
\\
\indent \citet{Koester:2007p484,Koester:2007p485} follow \citet{Gladders:2000p360} and \citet{Annis:1999AAS...195.1202A} using a technique known as maxBCG to search directly for over-densities of red sequence galaxies. They take advantage of the uniform colours and luminosities of BCGs in the selection procedure. The likelihood that each galaxy in a photometric sample has the photometric properties of a BCG and resides in an over-density of red sequence galaxies is evaluated at a grid of assumed redshifts. The redshift which maximises the likelihood (hence the name maxBCG) is used as a first estimate of the cluster redshift. Using this centre, all red sequence galaxies within  $1$ h$^{-1}$ Mpc projected radius are potential cluster members.  A percolation technique is then used in order to determine if the galaxy in question is in fact the cluster centre. The catalogue sample we employed contains $12,766$ clusters with photometric redshifts between $0.1$ and $0.3$ and is approximately volume limited. Koester at al. use the maxBCG algorithm on realistic mock catalogues and find that the sample is more than $90\%$ pure and more than $85\%$ complete for clusters with masses $\geq 1\times10^{14}\;M_{\odot}$. We extract BCG positions (which are defined to be at the cluster centres in this catalogue) and redshifts in order to select red sequence galaxies from the SDSS. Koester et al. find an rms scatter of only $\sigma = 0.015$ between the clusters' photometric redshift estimates and spectroscopic redshifts of the BCG for the 7813 clusters in their catalogue with spectroscopy in the SDSS.
\\
\indent \citet{Dong:2008p220} use a variant of the matched filter technique previously used by \citet{Postman:1996AJ....111..615P,Kawasaki:1998A&AS..130..567K,Schuecker:1998p521,Kepner:1999ApJ...517...78K,Bramel:2000p520,Kim:2002p370}, among others, which fits the distribution of galaxies in magnitude and position space to a standard spatial profile \citep[such as that predicted by][]{Navarro:1996p399}, using prior knowledge of spectroscopic or photometric galaxy redshifts and the galaxy luminosity function. The technique does not explicitly fit for the red sequence to select galaxies and hence can identify clusters of blue galaxies (if they exist). The algorithm generates a cluster likelihood map in position-redshift space whose peaks correspond to positions where the matches between the survey data and the cluster filter are optimised. Using a realistic mock SDSS catalogue Dong et al. show that the catalogue is $\approx 85\%$ complete and over $90\%$ pure for clusters with masses above $1.0 \times 10^{14}h^{-1} M_{\odot} $ and redshifts up to $z=0.45$. The errors for estimated cluster redshifts are typically less than $0.01$ (comparable to the photometric redshift errors of individual red-sequence galaxies) . The positions of clusters in the catalogue are the geometric centres of the galaxy distributions and do not necessarily correspond to a single galaxy position. Dong et al. do not identify BCGs in their catalogue. The matched filter has a faint end absolute magnitude cut off at $0.4L_\star$ and hence the catalogue is approximately volume limited. The value of $L_\star$ used assumes passive evolution, which may not be valid in detail.

\subsection{Cluster Galaxy Selection from SDSS}
\begin{figure}
	\centering
	\includegraphics[width=3in]{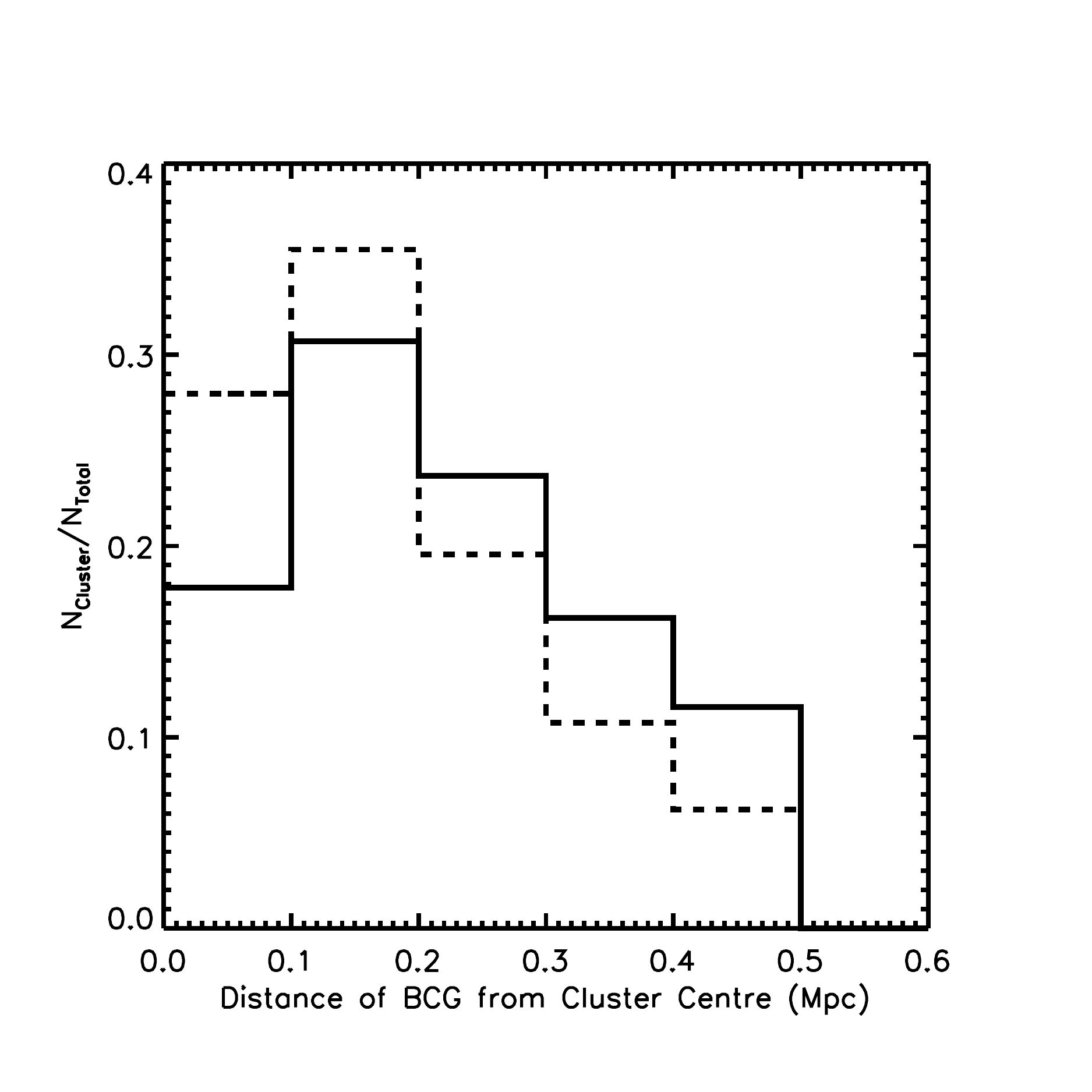}
	\caption{Histogram of projected distances of the BCG from the cluster centre. The Dong et al. sample is shown in the solid line and the Koester et al. sample in the dashed line. In the Dong et al. sample we find that $80\%$ of galaxies are within $350$ kpc and $50\%$ are within $200$ kpc. In the Koester et al. sample we find $90\%$ within $350$ kpc and $63\%$ within $200$ kpc.}
	\label{fig:disthisto}
\end{figure}

\indent In this study we use both catalogues but will concentrate on the results from the Dong et al. matched filter catalogue due to its larger size and deeper redshift coverage. In the following discussion, unless stated otherwise the values quoted and figures shown refer to the Dong et al. catalogue. In order to study the alignment of the BCG with the cluster we require the cluster centre and redshift in order to identify member galaxies. We define the cluster's shape using the identified member galaxies and identify the BCG as the brightest of these members.
\\
\indent To construct our sample of cluster galaxies we select galaxies from the SDSS database within $1$ Mpc projected of a given search centre (the geometric cluster centre quoted in the catalogue). We select only red sequence galaxies at the cluster redshift provided in the catalogue, using red sequence colour cuts $g-r$ and $r-i$ determined by \citet{Loh:2006p425}. The model magnitudes that we use are corrected for Galactic extinction using the maps by \citet{Schlegel:1998p1216}. We initially performed two colour cuts on the data corresponding to selecting objects at a given redshift that fall within the $2\sigma$ and $3\sigma$ ellipses in colour-colour space of the Loh \& Strauss red sequence. However, we found that our results do not strongly depend on the colour cut used, and hence we shall present only the data for the $3\sigma$ cut in this paper. The colour-colour ellipses used become larger at higher redshift to account for increasing photometric errors. One might suspect that this will also increase contamination by foreground and background galaxies and thus make the resultant clusters rounder. However, we find that our high redshift clusters are not systematically rounder than our low redshift clusters, suggesting that any redshift-dependent contamination is not severe. In addition to being more cleanly separated from the field galaxies in colour space, and therefore showing higher contrast and less contamination, the red sequence galaxies tend to be deeper in the potential well of their clusters, and therefore will do a better job of reflecting the true cluster shape.
\\
\indent After performing the colour cuts, the brightest galaxy in the $i$-band within the $1$ Mpc radius from the search centre is selected as the BCG. In $85\%$ of the clusters in common between the Dong et al. and Koester et al. catalogues, our identification of the BCG matched that of Koester et al. As BCGs are observed to be a highly uniform population in colour and luminosity \citep[e.g..][]{Postman:1995p237} and since the SDSS probes a uniform range of magnitudes at each redshift, we impose a cut of $i<i_{BCG}-3$ on the member red sequence galaxies to ensure that we sample each cluster approximately to equal depths. At the highest redshift in the Dong et al. catalogue ($z=0.44$) we sample the cluster down to $i=21.67$, where the SDSS is still quite complete \citep{2001AJ....122.1104Y}. 
\\
\indent The preliminary centre of each cluster is defined as the mean position of all selected red sequence galaxies. The final cluster galaxy sample consists of all red sequence galaxies within $0.5$ Mpc of this preliminary centre and the final geometric cluster centre is taken as the mean position of these galaxies. We find that the ellipticity of our clusters does not strongly depend on using a $1$ Mpc or $0.5$ Mpc radius to select cluster members, nor are rich and poor clusters affected differently. This is a result of determining cluster shapes using radius-weighted second moments, which we describe below.
\\
\indent Of the $36,785$ clusters in the Dong et al. catalogue we can extract data for $23,106$ clusters (the others being outside the redshift range for which \citet{Loh:2006p425} empirically define the red sequence). We discard clusters in which the geometric centre is farther than $0.5$ Mpc from the original search centre (this affects $1395$ clusters) as well as those clusters in which the BCG is farther than $0.5$ Mpc from the geometric centre ($7658$ of the remaining clusters). In the maxBCG routine of Koester et al. this BCG-miscentering may result from the algorithm itself. In a cluster with no clearly dominant BCG, the choice of BCG by the cluster-finder is less well-defined. Since this routine defines the BCG as the cluster centre, by removing clusters with clearly offcentre BCGs we may in fact be removing clusters with less dominant BCGs. However, since our alignment results from the Koester et al. catalogue match those of the Dong et al. catalogue (which does not depend on the BCG to determine cluster centres) it does not seem likely that such a systematic effect would bias our sample towards clusters with more dominant BCGs. 
\\
\indent A candidate BCG that is far from the centre may in fact simply be a bright foreground galaxy not physically associated with the cluster, or may indicate that the system is a pair of merging clusters. In the latter case there may exist two BCGs that may have reflected the shape of their original parent cluster but not necessarily of the merged cluster. Figure \ref{fig:disthisto} shows the histogram of projected distances of the BCG from the cluster centre. The BCG is within $350$ kpc of the centre in $80\%$ of the clusters in the Dong et al. sample and in $90\%$ of the clusters in the Koester et al. sample. Similarly \citet{Postman:1995p237} find that for $90\%$ of the clusters in their $z<0.05$ sample the BCGs lie within $350$ kpc. \citet{Skibba2010AAS...21533002S}, using a galaxy group catalogue constructed from the SDSS, find that the brightest galaxy does not reside at the halo centre (i.e., the position of lowest specific potential energy) in $25\%-40\%$ of their sample. The cuts imposed on the data for both the Dong et al. and Koester et al. catalogues are described in Table \ref{tab:cutsummary}. 

\begin{table*}
\centering
\caption{Cluster Samples}
\begin{tabular}{@{}lcc@{}}
\hline
\null&Dong et al. & Koester et al.\\
\hline
Sky Coverage &$\approx 6500$deg$^2$ &$\approx 7500$deg$^2$ \\
Total Input Catalogue & 36785 & 12766 \\
$0.08<z<0.44$ & 23106 & 12766\\
Centre$_{Geometric}$ $< 0.5$ Mpc from Centre$_{Input}$ & 21711 & 12202\\
BCG $ < 0.5$ Mpc from Centre$_{Geometric}$ & 14053 & 10754\\
\hline
$\chi^2 > 2.3$ & 9115 & 7071\\
$N_{Members}>5$ & 8081 & 6626 \\
$b/a_{BCG}<0.9$ & 7031 & 5744\\
Centre (BCG $\leq0.2$ Mpc from Centre$_{Geometric}$)& 3618 & 3647 \\
Offset (BCG $>0.2$ Mpc from Centre$_{Geometric}$) & 3413 & 2097 \\
\end{tabular}
\label{tab:cutsummary}
\end{table*}

\indent To compare the two catalogues we have matched cluster centres, requiring that their projected distance be less than $2$ Mpc and their redshifts agree to within $0.02$. The sample of clusters we compare is restricted to the redshift range of the Koester catalogue ($0.1<z<0.3$) which corresponds to $3189$ clusters from the Dong et al. sample and $6625$ clusters from the Koester et al. sample. Our matching routine returns a total of $1000$ matching clusters between the two samples, heavily weighted to the richer systems. The geometric centres agree (separation less than $300$kpc) in $990$ cases and  the selected BCGs match in $984$ cases. For clusters that do not have matching BCGs, the initial search centres are on average more than $300$ kpc apart. The richness distributions are very similar in both catalogues.

\section{Analysis}
\subsection{Cluster Shapes}
\indent{We follow the method used by \citet{Kim:2002ASPC..268..395K} to calculate the orientation and elongation of clusters. We first calculate the radius-weighted second moments (the first moments vanish by definition of the cluster centre) for all cluster members:}
\\
\begin{equation}
M_{xx}\equiv\;\left\langle\;\frac{x^{2}}{r^2}\right\rangle\;,\;M_{xy}\equiv\;\left\langle\;\frac{xy}{r^2}\right\rangle\;,\;M_{yy}\equiv\;\left\langle\;\frac{y^{2}}{r^2}\right\rangle
\label{eqn:moments}
\end{equation}
where $x$ and $y$ are the respective distances of a given member galaxy from the cluster centre defined in the equatorial coordinate system. Using the relation between the Stokes parameters $Q$ and $U$, the position angle ($\phi$) and axis ratio ($\alpha\equiv\frac{b}{a}$, where $a$ and $b$ are the semi-major and semi-minor axis of the ellipse) we may determine $\phi$ and $\alpha$ for each cluster.  In order to compare the position angle of the clusters to those of galaxies reported in the SDSS database we convert $\phi$ from West of North to East of North.
\\
\begin{gather}
Q\equiv\frac{1-\alpha}{1+\alpha}\cos(2\phi)=M_{xx}-M_{yy}=2M_{xx}-1 \\
U\equiv\frac{1-\alpha}{1+\alpha}\sin(2\phi)=2M_{xy}
\end{gather}
{The uncertainties in $Q$ and $U$ are given by Poisson statistics as:}
\begin{gather}
\sigma_Q\equiv\biggl[\frac{2}{N(N-1)}\Sigma\biggl(\frac{x^{2}}{r^2}-\left\langle\frac{x^{2}}{r^2}\right\rangle\biggr)^2\biggr]^{\frac{1}{2}} \\
\sigma_U\equiv\biggl[\frac{2}{N(N-1)}\Sigma\biggl(\frac{xy}{r^2}-\left\langle\frac{xy}{r^2}\right\rangle\biggr)^2\biggr]^{\frac{1}{2}}
\end{gather}
\\
 From the above it is straightforward to solve for $\phi$ and $\alpha$:
 \begin{equation}
\phi=\frac{1}{2}\arctan\biggl(\frac{U}{Q}\biggr)\;,\;\alpha=\frac{1-D}{1+D}\;,\;D\equiv\sqrt{Q^2 + U^2}
\label{eqn:phi}
\end{equation}

\begin{figure}
	\centering
	\includegraphics[width=0.5 \textwidth]{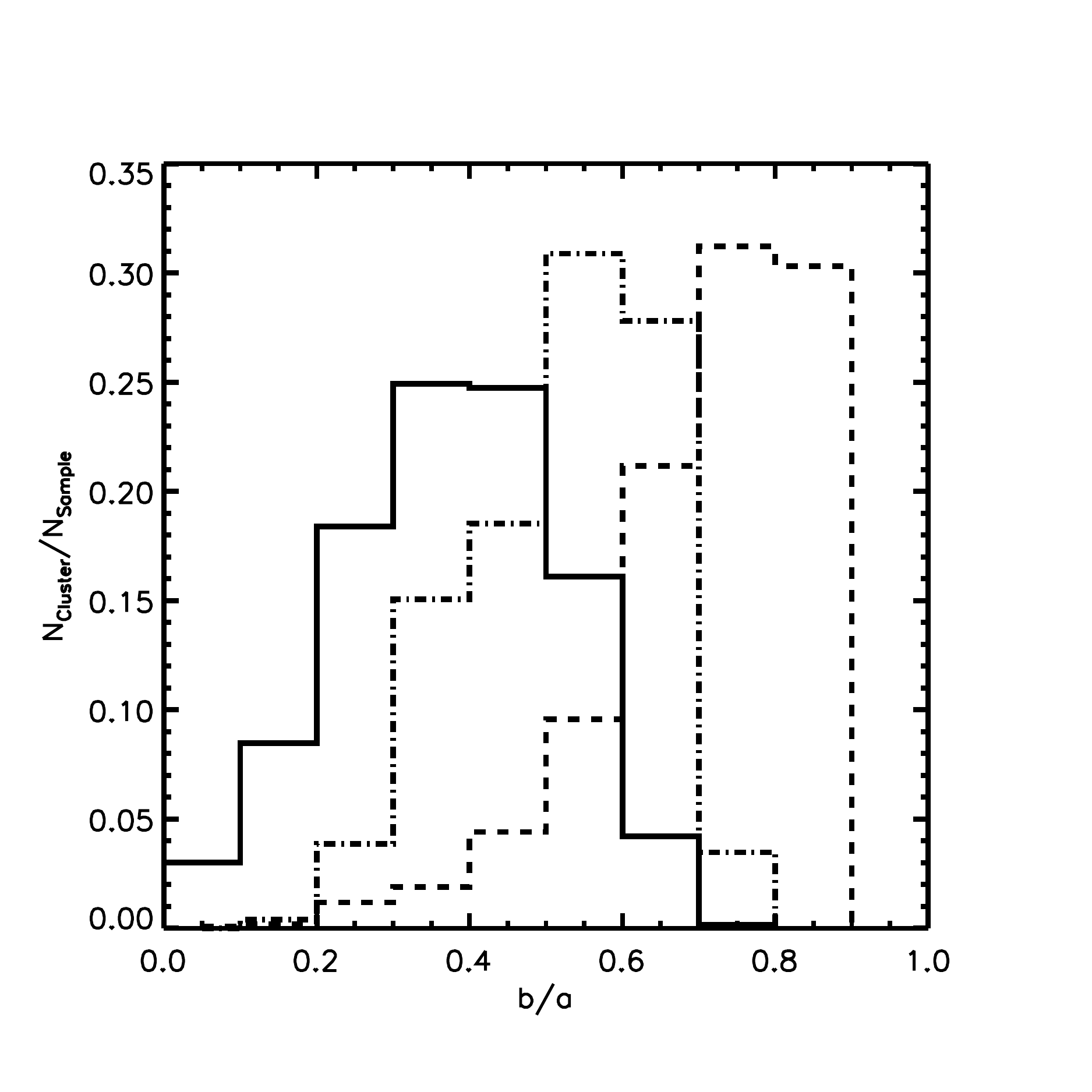}
	\caption{Distribution of axis ratios $b/a$ of clusters and BCGs.  The solid line shows the ellipticities of clusters, as determined by our algorithm. The dash-dotted line shows the ellipticities of the $259$ clusters with more than $20$ members. In these clusters the shape should be determined with higher signal-to-noise than in the poorer clusters. They are less elliptical than the total sample of clusters. The difference results from the fact that sparse sampling in the poor clusters has a tendency to make them more elliptical \citep[see Appendix A in][]{Allgood2006MNRAS.367.1781A} than more highly sampled clusters. The dashed line shows the distribution of axis ratios for BCGs given by the SDSS database. In our analysis we remove BCGs with $b/a>0.9$ for which the position angle determination is meaningless.}
	\label{fig:abhisto}
\end{figure}

\begin{figure}
	\centering
	\includegraphics[width=0.5 \textwidth]{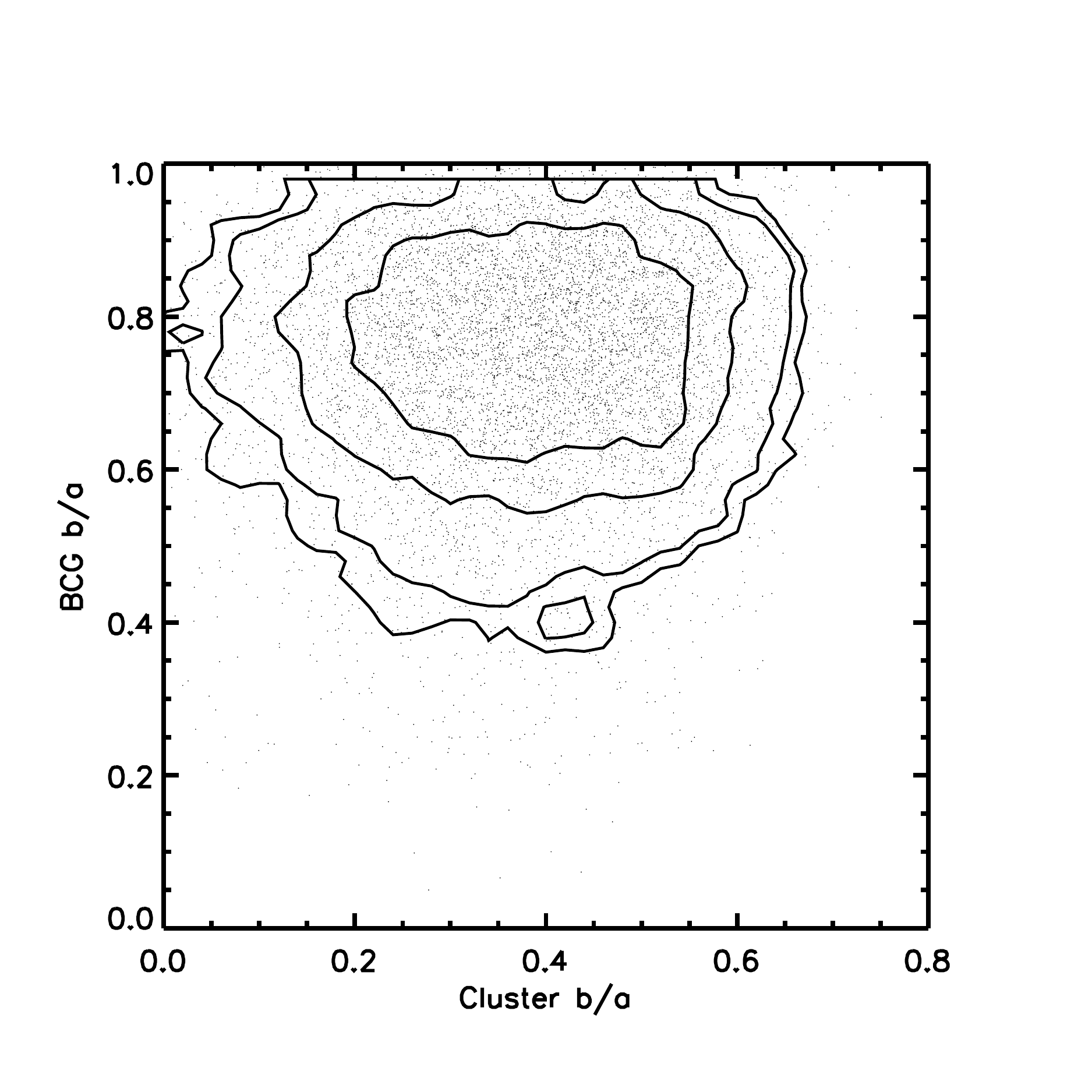}
	\caption{Comparison of the axis ratio $b/a$ of BCGs and clusters. There is no correlation between the two axis ratios. }
	\label{fig:abcomp}
\end{figure}

\begin{figure}
	\centering
	\includegraphics[width=0.5 \textwidth]{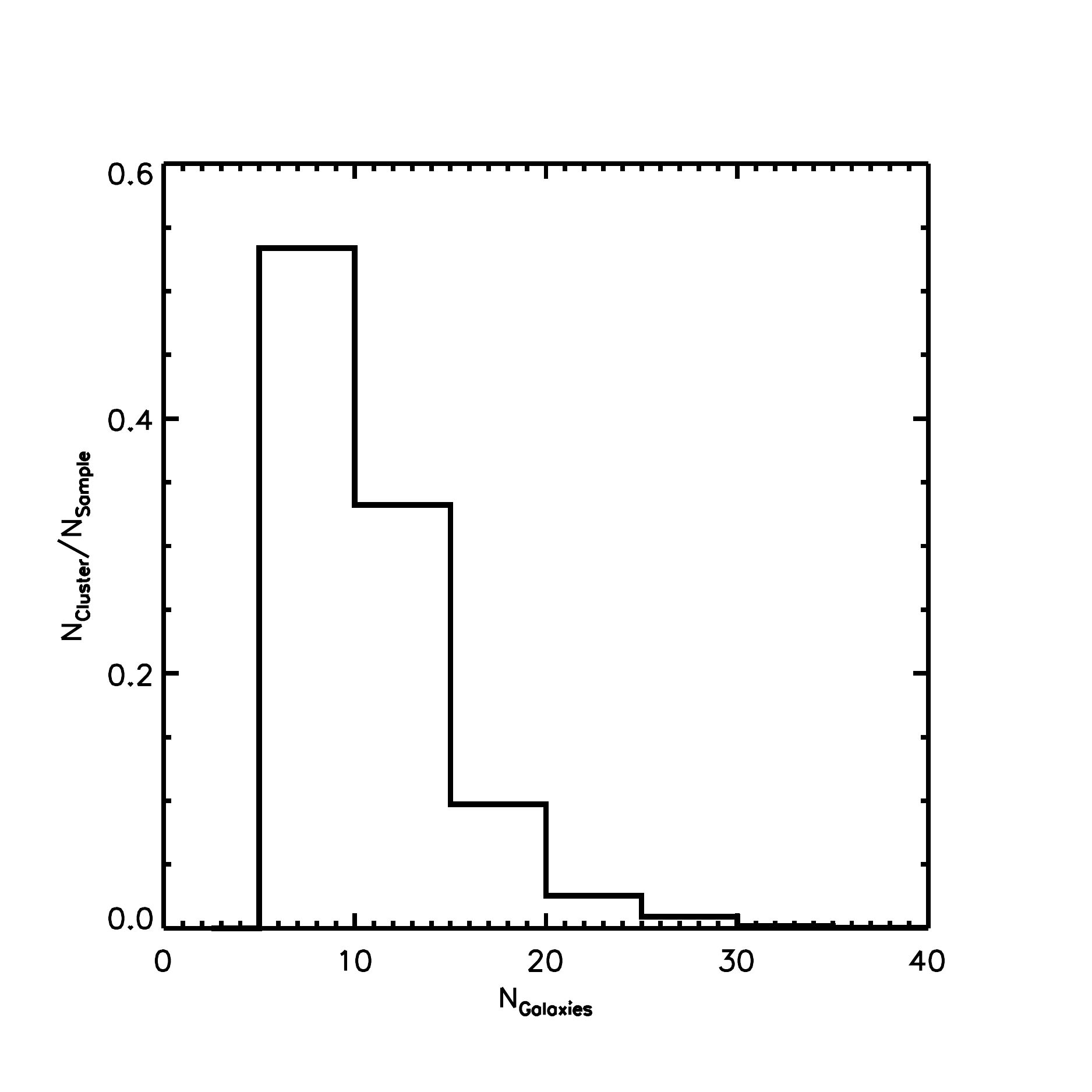}
	\caption{Distribution of member counts in clusters (Total sample), which is used as a cluster-richness measure in the following analysis.  The count includes red sequence galaxies up to $3$ magnitudes fainter than the BCG.}
	\label{fig:richhisto}
\end{figure}

\noindent The position angle and ellipticity of the BCG are provided in the SDSS data, based on a two-dimensional fit to a PSF-convolved de Vaucouleurs profile \citep{Stoughton:2002p359}.
\\
\indent The cluster and BCG position angles are distributed isotropically, as is expected. Clusters are elliptical rather than round (Figure \ref{fig:abhisto}) which is in agreement with observations by e.g.. \citet{Carter:1980p229} and \citet{Binggeli:1982p376} and supported by simulations \citep[e.g..][]{Splinter:1997p81,Hopkins:2005p348}. Our population of clusters is more elliptical than those of the Hopkins et al. simulations (who find a mean ellipticity of $b/a \approx 0.7$ at $z=0$), with the richer clusters having a mean ellipticity closer to, but still more elliptical than, the values predicted by the simulation (Figure \ref{fig:abhisto}). The shapes determined by Hopkins et al. are calculated using the entire cluster (all particles within the friends-of-friends radius, which is greater than $0.5$ Mpc in all cases), whereas we employ only the central $0.5$ Mpc. Hopkins et al. note that the measured ellipticities decrease with decreasing radius, however, \citet{Allgood2006MNRAS.367.1781A} have noted that the central parts of dark matter halos are more aspherical than the overall cluster. It is possible that this effect is the cause of the difference in the values of ellipticity determined by us and those measured in the Hopkins et al. simulations. There is no correlation between the cluster and BCG ellipticities (Figure \ref{fig:abcomp}), that is highly elongated BCGs show no tendency to live in highly elongated clusters. \citet{Hashimoto:2008MNRAS.390.1562H} found the same result measuring the ellipticities of x-ray emission of clusters. 
\\
\indent Clusters with offset BCGs are potentially two merging clusters and the shape of the selected BCG may be related to the shape of its original host cluster,rather than the galaxy distribution of two merging clusters. Hence we might expect a weaker alignment between these clusters and their BCGs. We divide the sample roughly in two by splitting at a BCG-centre distance of $0.2$ \rm{Mpc} and we call these the Centre and Offset samples respectively. Figure \ref{fig:richhisto} shows the distribution of the number of red sequence galaxies for the final cluster sample. The Offset sample has slightly poorer clusters on average than does the Centre sample.
\\
\indent Clusters with very small axis ratios (i.e. very elongated) are somewhat unusual objects. Sixty percent of the $212$ clusters in our final sample with $b/a<0.1$ have offset BCGs (vs. $50\%$ in the total sample), and $94\%$ have fewer than 8 members. These objects may be less virialized, and are certainly less rich systems. We find that excluding them does not qualitatively change our statistics or conclusions.
\\
\indent  We remove clusters that are consistent with being round ($4938$ clusters), as we will be unable to measure an alignment signal for such objects.  Following \citet{Kim:2002ASPC..268..395K} we define this condition as:
\\
\begin{equation}
\chi^2\equiv\biggl(\frac{Q}{\sigma_Q}\biggr)^2+\biggl(\frac{U}{\sigma_U}\biggr)^2\leq 2.3
\label{eqn:chisq}
\end{equation}
\\
which marks $68\%$ in a $\chi^2$ distribution with two degrees of freedom i.e. we require clusters to be farther than $1\sigma$ away from $D=0$ (see Equation \ref{eqn:phi}) which defines a round cluster. We also remove the $1034$ remaining clusters that have four or fewer members on the red sequence, as a position angle defined from them will be dominated by shot noise. We remove from our final catalogue those clusters whose BCGs are round, which we define as $b/a>0.9$ since their position angles will be meaningless. This eliminates an additional $1050$ clusters. The final cluster catalogues are summarised in second part of Table \ref{tab:cutsummary}. Throughout the subsequent analysis we use the cluster sample labelled as $b/a_{BCG}<0.9$ in Table \ref{tab:cutsummary} as the Total sample and split this into the Centre and Offset samples.

\subsection{Alignment}
\begin{figure}
	\centering
	\includegraphics[width=0.5 \textwidth]{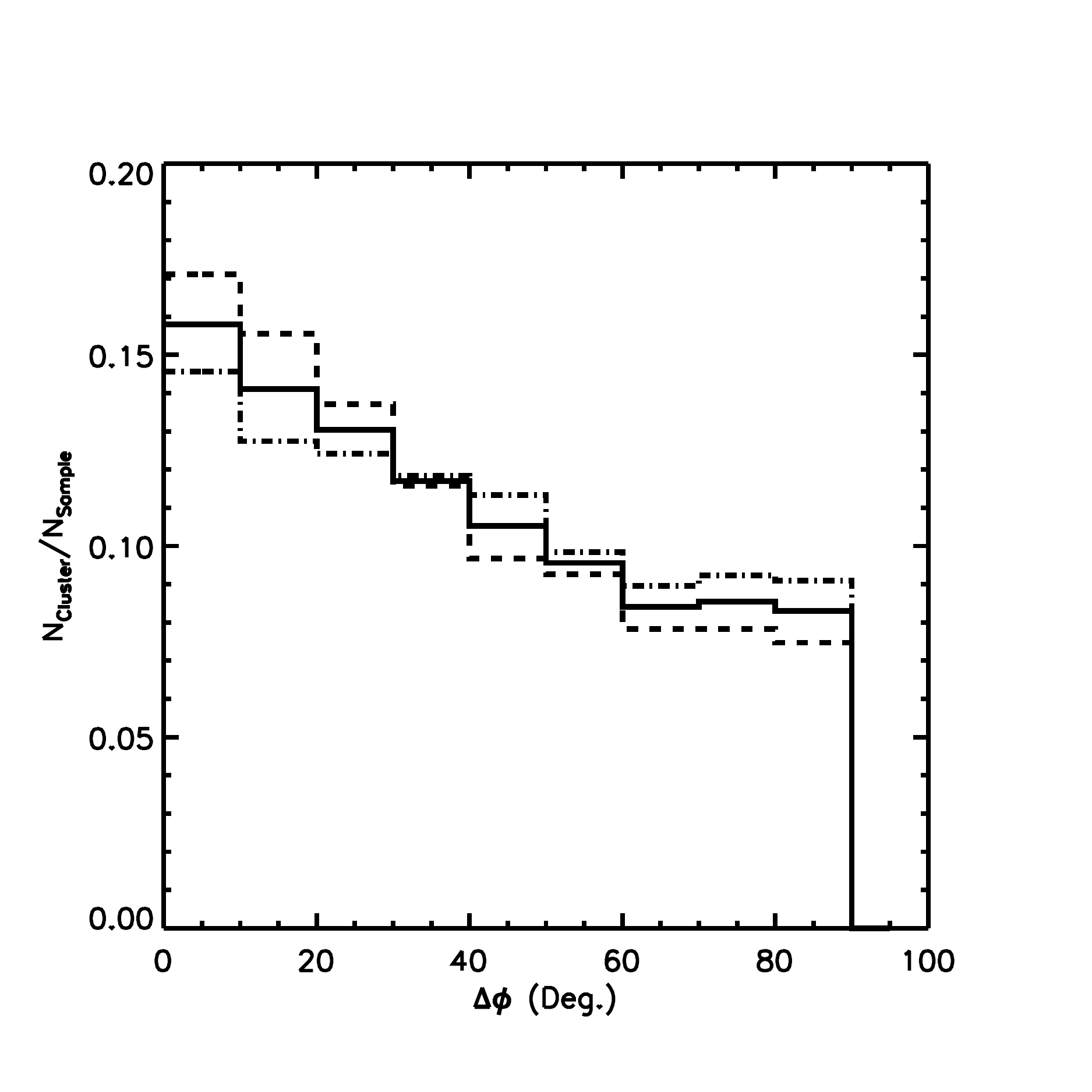}
	\caption{Histogram of cluster-BCG alignment in the Total (solid line), Centre (dashed line) and Offset(dash-dotted line) samples.  The distributions are not flat. The alignment signal for the Centre sample is approximately $5.5\sigma$ stronger than in the Offset sample.}
	\label{fig:histocuty}
\end{figure}

\begin{figure}
	\centering
	\includegraphics[width=0.5\textwidth]{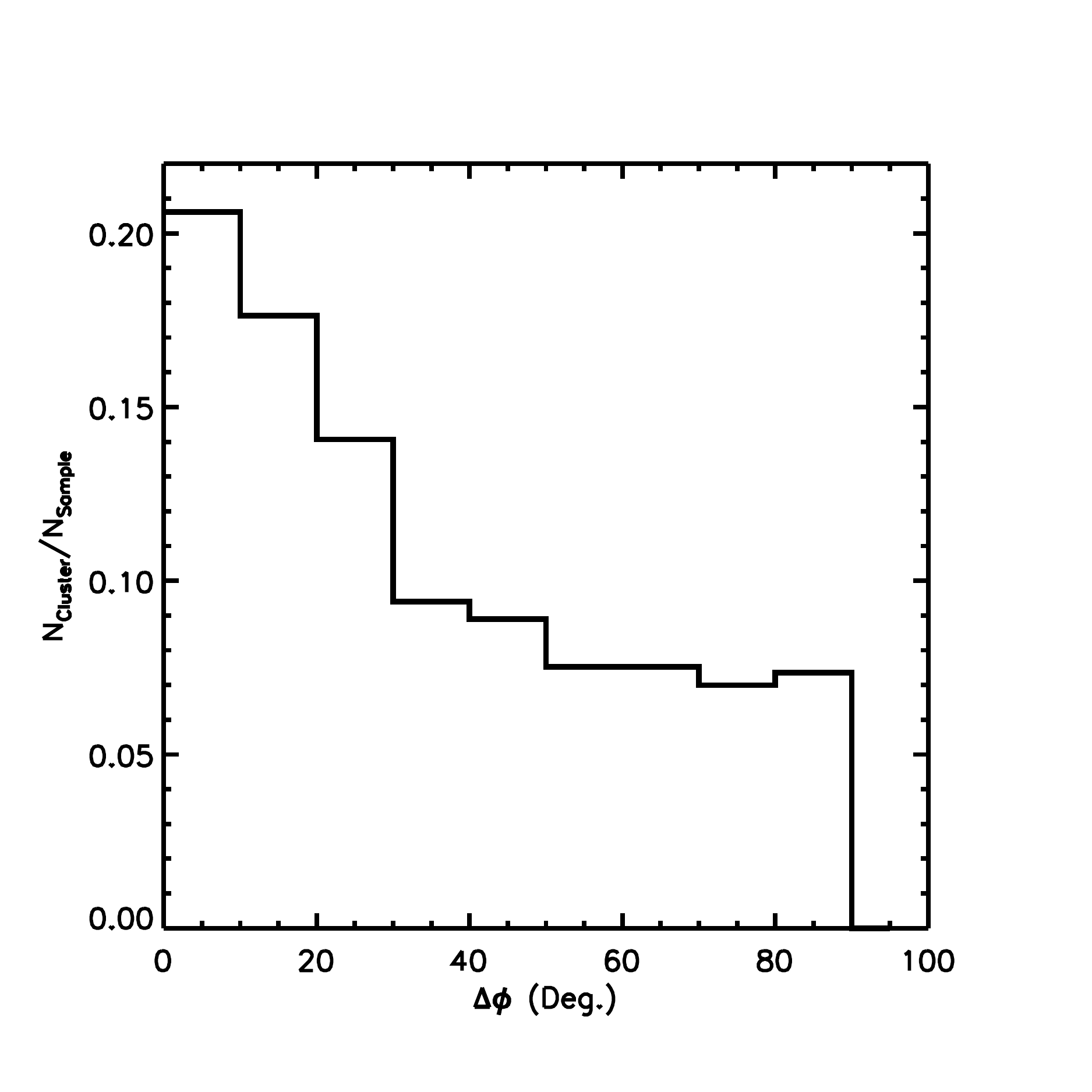}
	\caption{Histogram of alignment between the cluster and the BCG-centreline. The centreline is the line drawn from the cluster centre to the BCG. This test is only performed on the Offset sample whose BCGs are farther than $0.2\;Mpc$ from the centre. We find a very strong alignment signal, with $\mathcal{R}$$=1.10\pm0.04$.}
	\label{fig:centerphi}
\end{figure}

\indent With our measurements of the position angles of clusters in hand, we investigated a number of alignment signals between BCGs and their host clusters. We define the alignment of the BCG with the cluster as

\begin{equation}
\Delta\phi\equiv\vert\phi_{BCG}-\phi_{Cluster}\vert
\end{equation}

and we say that the BCG and cluster are aligned for $\Delta\phi\leq30^{\circ}$, following \citet{Binggeli:1982p376}.
\\
\indent We quantify the strength of the alignment signal by calculating the ratio $\mathcal{R}$ of the number of clusters with $\Delta\phi \le 30^{\circ}$ to the number with $\Delta\phi > 30^{\circ}$. This definition of the alignment signal gives an indication of the steepness of the distribution as well as the relative number of aligned and non-aligned cluster-BCG pairs. The errors are determined assuming that the counts are Poisson processes with uncertainty $\sigma_a=\sqrt{a}$ where $a$ is the count, and that the uncertainties in the two counts are independent. A random distribution would yield $\mathcal{R}$$=0.5$. We show the distribution of alignments in Figure \ref{fig:histocuty} and tabulate $\mathcal{R}$ in Table \ref{tab:lambdasumgen}. There is a clear and unambiguous alignment signal (the Binggeli effect). In the Centre sample the alignment signal ($\mathcal{R}$$=0.865$) is approximately $5.5\sigma$ stronger than in the Offset sample ($\mathcal{R}$$=0.659$) where the significance is defined as the difference between $\mathcal{R}_{Centre}$ and $\mathcal{R}_{Offset}$ weighted by the uncertainties summed in quadrature. We find only a substantially weaker alignment for second ranked galaxies ($\mathcal{R}$$=0.651$ for centred second ranked galaxies, $\mathcal{R}$$=0.533$ for offset) and random alignment for third ranked galaxies. The observed alignment is unique to the BCG which is another characteristic setting it apart from other cluster galaxies.
\\
\indent In the Offset clusters the line connecting the cluster centre and the BCG is strongly aligned with the cluster shape, with $\mathcal{R}$$=1.10\pm0.04$ (Figure \ref{fig:centerphi}). In a bottom-up structure formation scenario with galaxies moving in a cluster that is oriented along a filament, a line drawn from a massive galaxy to the centre of the cluster will be strongly aligned with the direction of elongation of the cluster. However, we may instead be seeing a purely geometrical effect: in an elongated cluster more galaxies lie near the major axis of the cluster than the minor axis, hence a line drawn from most galaxies to the centre will be aligned with the cluster. To investigate this we also look at the alignment for the fainter red sequence galaxies farther than $200$ kpc from the centre. If we are really seeing a purely geometric effect the alignment should be similar in these samples as well. We find that the samples have $\mathcal{R}$$=0.94, 0.92, 0.92, 0.86$ for the second, third, fourth, and fifth ranked red sequence galaxies respectively. The alignment in the second ranked etc. galaxies is significantly non-random even if it is roughly $3\sigma$ weaker than for the BCG. This suggests that the observed alignment is the result of a combination of influences. For the BCG, the effect of movement along the filament is stronger than for the other galaxies, which is why $\mathcal{R}$ is higher. However, since all fainter galaxies show a similar alignment the geometry of the cluster is likely to have a significant influence as well.

\begin{table*}
\centering
\caption{Alignment between BCG and host cluster.}
\begin{tabular}{@{}llcclcc@{}}
\hline
\null & \null& Dong et al.& \null \\
Sample & $N_{Cluster}$ & $\mathcal{R}$ & $\pm$ \\
\hline
Total& 7031&0.753&0.018  \\
Centre &3618& 0.865&0.030 \\
Offset &3413&0.659&0.022 \\
\hline
\hline
\null & \null & Dong et al. & \null & \null & \null  \\
\null & \null & ($0.1<z<0.3$) & \null & \null & Koester et al.  \\
Sample & $N_{Cluster}$ & $\mathcal{R}$ & $\pm$ & $N_{Cluster}$ & $\mathcal{R}$ & $\pm$ \\
\hline
Total &2719&0.833&0.032& 5744 & 0.858 & 0.023 \\
Centre &1365&0.975&0.052& 3647 & 1.00 & 0.033 \\
Offset &1354&0.710&0.039& 2097 & 0.649 & 0.029 \\
\end{tabular}
\label{tab:lambdasumgen}
\end{table*}

\subsection{The Effect of BCG Dominance}
\begin{figure}
	\centering
	\includegraphics[width=0.5\textwidth]{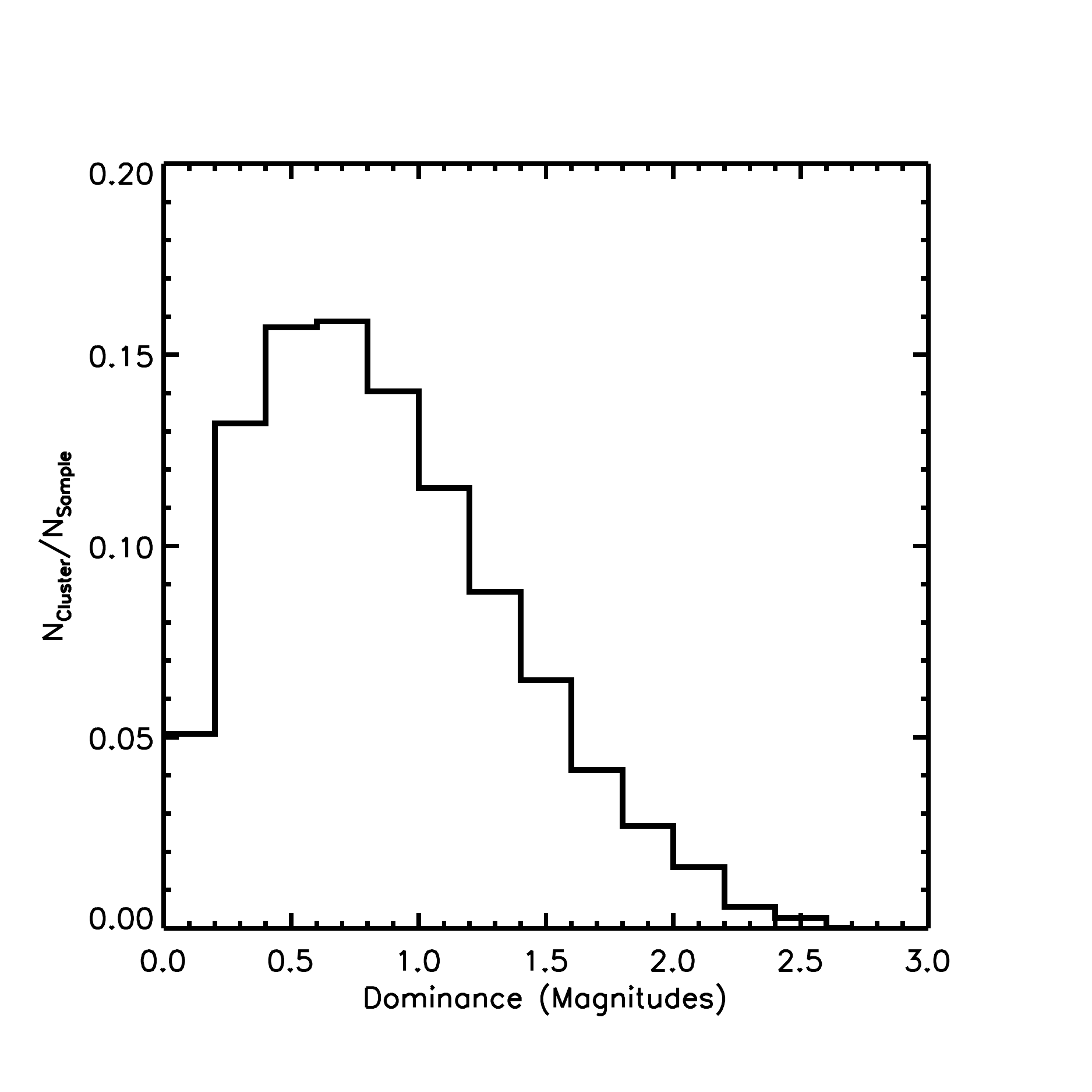}
	\caption{Histogram of BCG dominance over the other cluster galaxies. Dominance is defined as the difference between the $i$-band magnitude of the BCG and the average of the second and third ranked galaxies. In the analysis of the alignment effect, we split the sample into dominant (dom$\geq 0.65$) and non-dominant BCGs.}
	\label{fig:domhisto}
\end{figure}

\indent One readily observed distinguishing property of BCGs is their dominance in luminosity over other cluster galaxies. They do not simply form the statistical extreme of the cluster galaxy luminosity function but require a distinct physical formation process from other cluster galaxies. A statistical test developed by \citet{Tremaine:1977p403} and used by, among others, \citet{Loh:2006p425} has been used to show this using the difference between first and second ranked galaxies in clusters. The processes by which BCGs become luminous, however, are not well understood. Simulations by \citet[e.g.][]{Dubinski:1998p366} predict that the BCG forms through mergers of several massive galaxies moving along a filament early in the cluster's history. Since the alignment effect may also result from a preferred infall direction of galaxies during cluster formation a connection between dominance and alignment is worth investigating. Figure \ref{fig:domhisto} shows the distribution of the BCG's dominance, in the Dong et al. sample, defined following \citet{Kim:2002ASPC..268..395K} as the difference in $i$-band magnitude of the BCG and the mean magnitude of the second and third ranked galaxy. 

\begin{equation}
dom\equiv m_1-\frac{m_2+m_3}{2}
\end{equation}

\begin{figure*}
	\centering
	\subfigure[]{
	\includegraphics[width=0.49\textwidth]{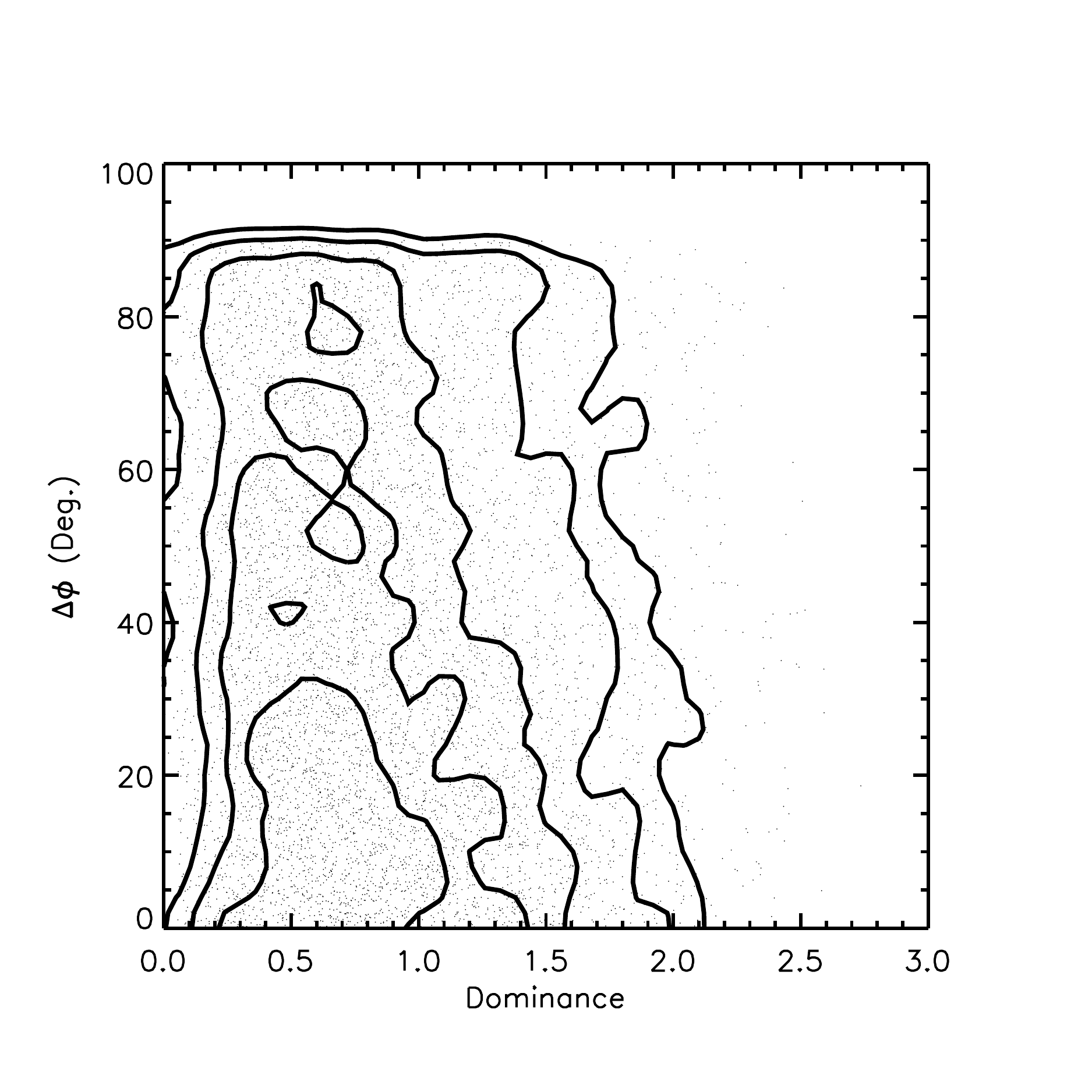}}
	\subfigure[]{
	\includegraphics[width=0.49\textwidth]{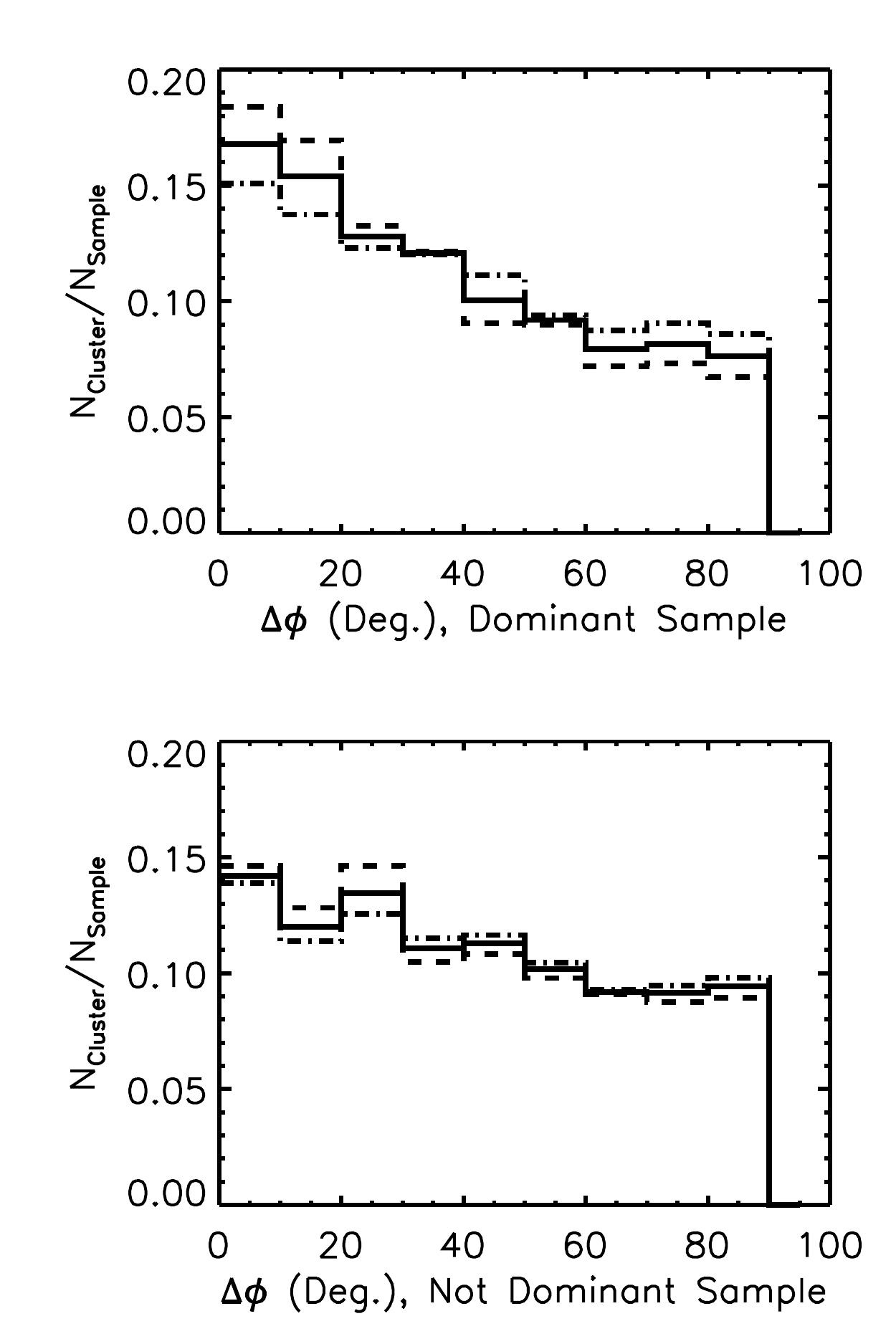}}
	\caption{Panel (a): Scatter plot of dominance vs. alignment. There is a clear indication that dominant clusters tend to be more aligned than non-dominant clusters. Panel (b): Histogram of cluster-BCG alignment examining the effect of BCG dominance on the alignment signal. The top panel shows the dominant sample (dom$\geq0.65$), and the bottom panel shows the non-dominant sample. The Total (solid line) and Centre (dashed line) samples have a significant difference between dominant and non-dominant clusters. The Offset sample shows a smaller, statistically less significant, difference between the two sub-samples. The $\mathcal{R}$ statistics for these various samples are given in Table \ref{tab:domphi}.}
	\label{fig:histodom}
\end{figure*}

An alternative would be to investigate the dependence of alignment on the absolute magnitude of the BCG. However, if the BCG grows by merging with the second and third ranked galaxies this measure of dominance is more appropriate. The dominance distribution is more strongly skewed than that of \citet{Kim:2002ASPC..268..395K}, presumably because of the differences in the details of the cluster selection algorithms. The mean dominance for the Centre sample is $0.90$ mag and the mean dominance for the Offset sample is $0.81$ mag. \citet{Loh:2006p425} find a magnitude difference between first and second ranked galaxies of $~0.8$ mag, which is in good agreement with our values. 
\\
\indent If the BCG's high luminosity and its alignment with their host cluster stem from infall and merging along a preferred direction one might expect a correlation between BCG dominance and its degree of alignment with the cluster. \citet{Kim:2002ASPC..268..395K} found in a small SDSS sample that $39$ out of $66$ dominant BCGs show strong alignment ($\Delta \phi < 30^{\circ}$) but find no statistically significant alignment in a sample of $49$ non-dominant BCGs. In a scatter plot of dominance vs. alignment (Figure \ref{fig:histodom}a) we already see an indication of stronger alignment (smaller $\Delta \phi$) in high dominance clusters. Defining as dominant those BCGs that have dom$\ge 0.65$, which corresponds roughly to the peak of the dominance distribution (Figure \ref{fig:domhisto}) we find that the dominant sample has an approximately $4.4\sigma$ stronger alignment than the non-dominant sample (see Figure \ref{fig:histodom}b and Table \ref{tab:domphi} for $\mathcal{R}$ values). However, the non-dominant sample is not completely isotropic.  In the Centre cut the dominant sample is more strongly aligned than the non-dominant sample (approximately $3.7\sigma$). The difference between the dominant and non-dominant sub-sample of the Offset cut is significant at only the $2.1 \sigma$ level.
\\
\indent Using Kolmogorov-Smirnov tests we find with high confidence (Table \ref{tab:domks}) that the $\Delta \phi$ distribution of the total dominant sample is drawn from a different underlying distribution than the non-dominant sample. Similarly we find that the Centre dominant sample comes from a different underlying distribution than the non-dominant sample. In the Offset sample the difference between the dominant and non-dominant samples is less significant.

\begin{table*}
\centering
\caption{Alignment between BCG and clusters for dominant and non-dominant BCGs.}
\begin{tabular}{@{}llcclcc@{}}
\hline
\null & \null& Dong et al.& \null  \\
Sample & $N_{Cluster}$ & $\mathcal{R}$ & $\pm$  \\
\hline
Total Dominant & 4353&0.818&0.025 \\
Total Not-Dominant &2678&0.658&0.026  \\
Centre Dominant &2256&0.944&0.040 \\
Centre Not Dominant &1154& 0.728&0.043 \\
Offset Dominant&2097&0.700&0.031 \\
Offset Not Dominant&1520&0.608&0.032 \\
\hline
\hline
\null & \null & Dong et al. & \null & \null & \null  \\
\null & \null & ($0.1<z<0.3$) & \null & \null & Koester et al.  \\
Sample & $N_{Cluster}$ & $\mathcal{R}$ & $\pm$ & $N_{Cluster}$ & $\mathcal{R}$ & $\pm$  \\
\hline
Total Dominant & 1692& 0.921 & 0.045 & 4009 & 0.927 & 0.029 \\
Total Not-Dominant & 1025 & 0.705 & 0.045 & 1735 & 0.715 & 0.035 \\
Centre Dominant & 907 & 1.090 & 0.072 & 2722 & 1.065 & 0.041 \\
Centre Not Dominant & 457 & 0.778 & 0.073 &924 & 0.844& 0.058 \\
Offset Dominant& 785 & 0.756 & 0.055 &1287& 0.689 & 0.039 \\
Offset Not Dominant& 568 & 0.651 & 0.056 &793 &0.553 & 0.041\\
\end{tabular}
\label{tab:domphi}
\end{table*}

\begin{table}
\centering
\caption{Kolmogorov-Smirnov tests determining whether the dominant and non-dominant sub-samples are drawn from different underlying distributions. $\Delta$ gives the significance of the difference between $\mathcal{R}$ statistics for the dominant and non-dominant samples in units of standard deviation ($\sigma$).}
\begin{tabular}{@{}llcc@{}}
Sample & K-S Confidence &$\Delta$\\
\hline
Dong Total ($0.08<z<0.44$)
& $99.99\%$&4.4\\
Centre &$99.92\%$&3.7\\
Offset &$96.52\%$&2.1\\
\hline
Dong Total ($0.1<z<0.3$)
& $99.99\%$&3.4\\
Centre & $99.98\%$&3.0\\
Offset & $82.92\%$&1.3\\
\hline
Koester
Total & $99.99\%$&4.7\\
Centre & $99.76\%$&3.1\\
Offset & $91.39\%$&2.4 \\
\end{tabular}
\label{tab:domks}
\end{table}

\subsection{The Effect of Cluster Richness}
\begin{figure*}
	\centering
	\subfigure[]{
	\includegraphics[width=3in]{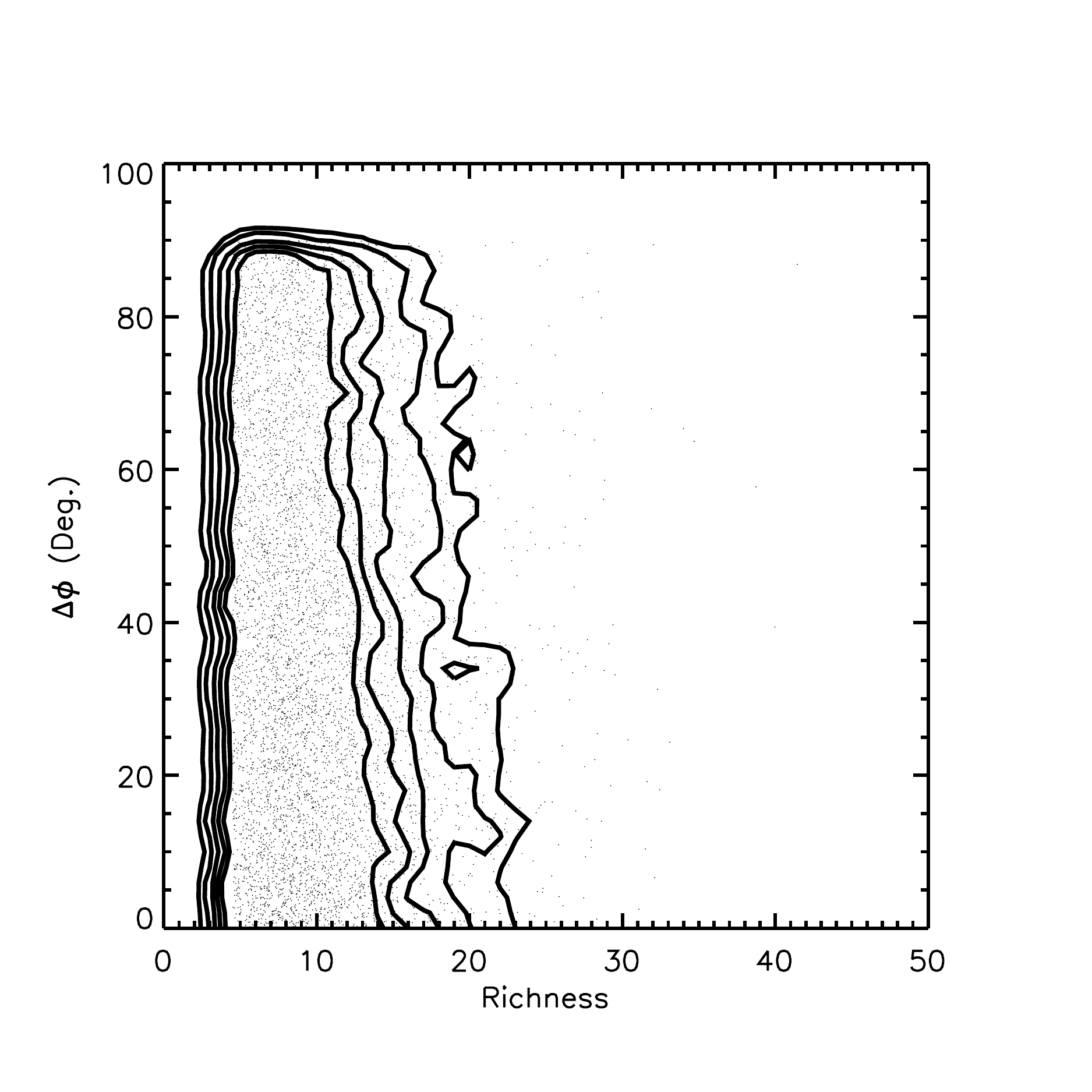}}
	\subfigure[]{
	\includegraphics[width=3in]{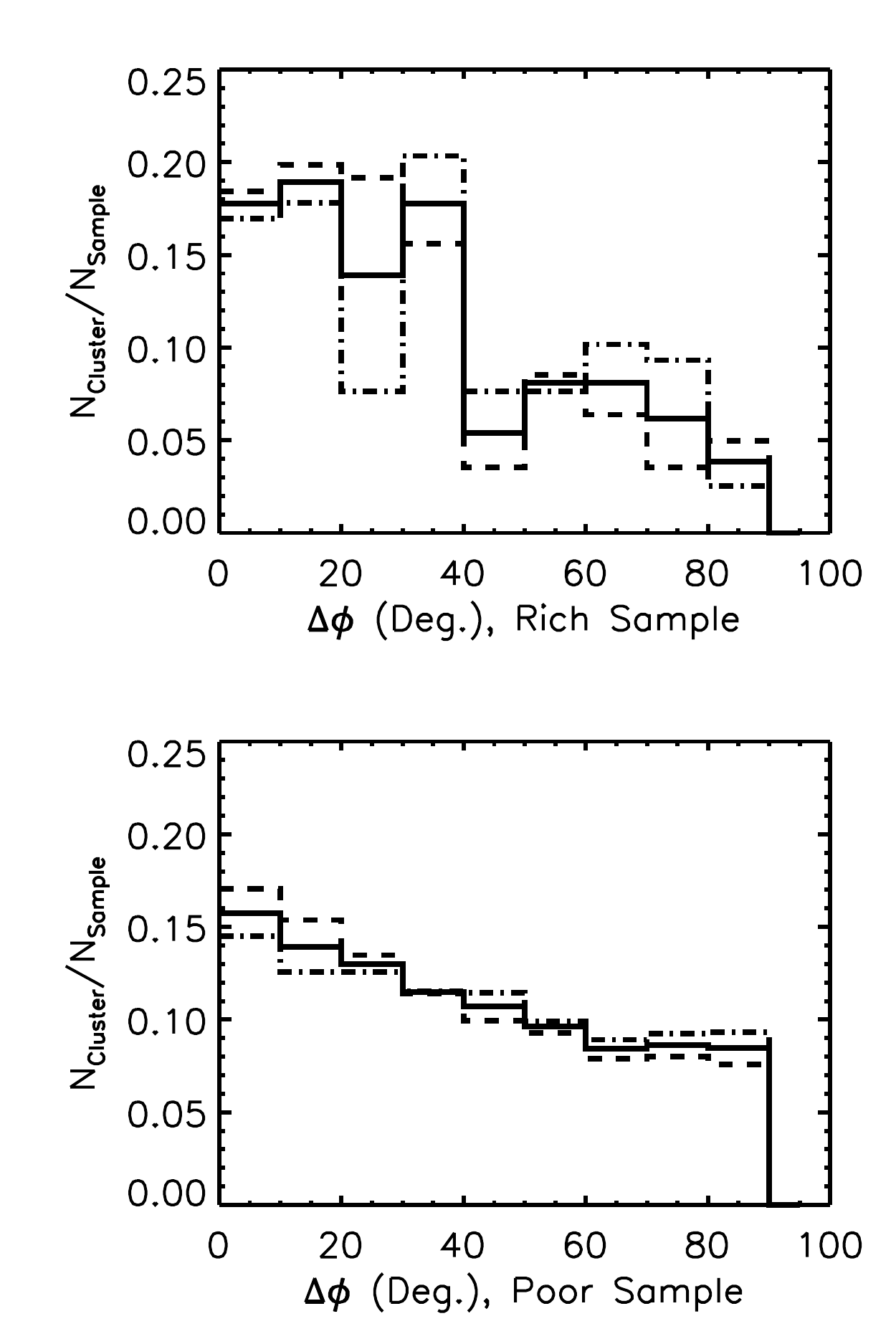}}
	\caption{Panel (a): Scatter plot of richness vs. alignment. Rich clusters show a stronger preference for alignment than poor clusters. Panel (b): Histogram of cluster-BCG alignment examining the effect of cluster richness on the alignment signal. The top panel shows the rich samples (Total sample solid line, Centre sample dashed line, Offset sample dash-dotted line) ($N_{Members}\geq20$), and the bottom panel shows the poor samples.  The alignment signal in the total rich sample is approximately $2.3\sigma$ stronger than in the poor sample (Table \ref{tab:richphi}, \ref{tab:ksrich}).}
	\label{fig:historich}
\end{figure*}

\begin{figure*}
	\centering
	\includegraphics[width=1\textwidth]{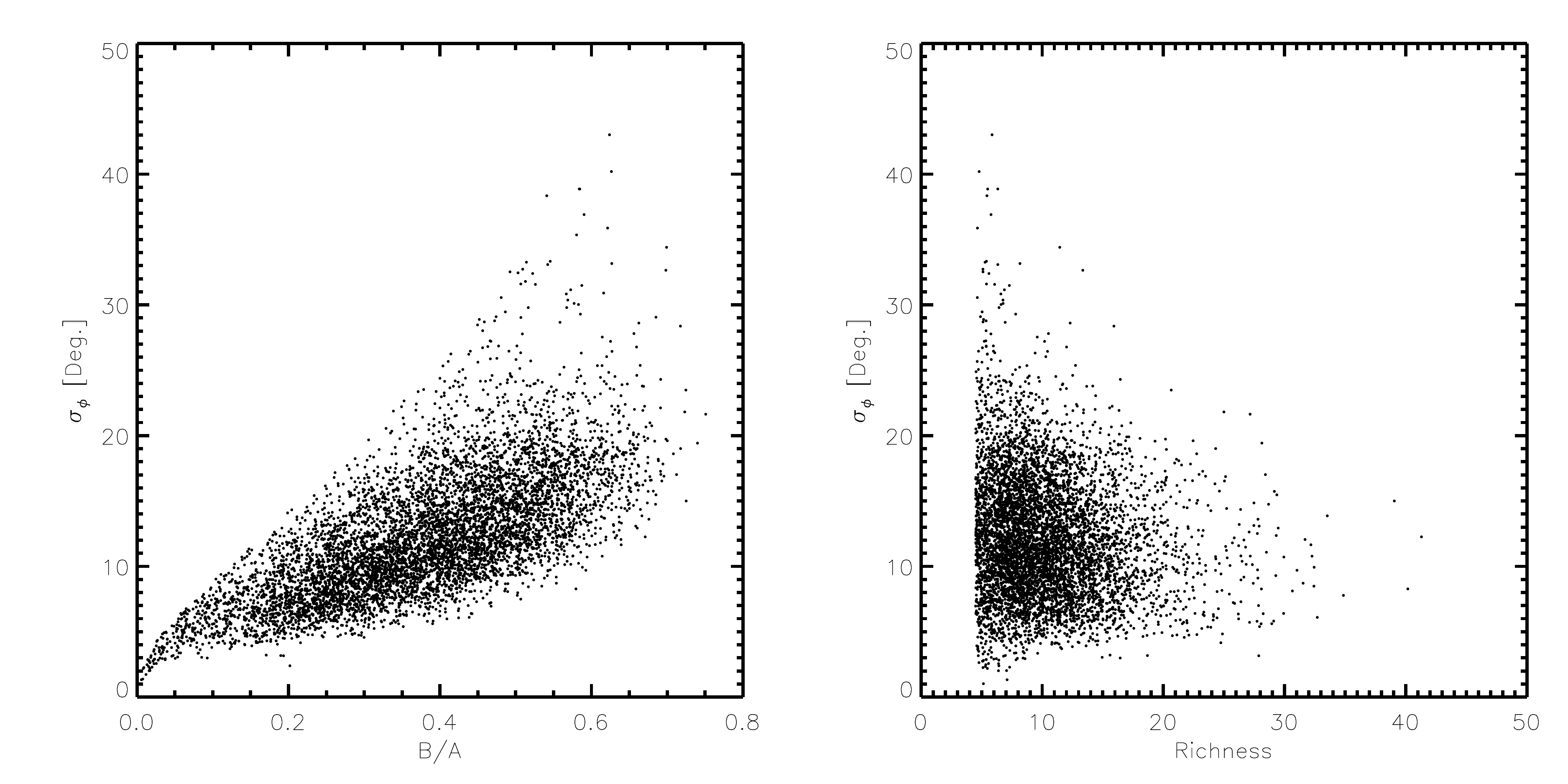}
	\caption{Left panel: Position angle uncertainty as a function of cluster axis ratio. The position angle uncertainty is determined by propagating the shot-noise errors on the Stokes parameters through Equation \ref{eqn:phi}. Right panel: Position angle uncertainty as a function of cluster richness. There is a clear trend towards higher position angle uncertainty in rounder clusters. The effect of the number of cluster members is much less clear.}
	\label{fig:uncert_ba_rich}
\end{figure*}

\begin{figure*}
	\centering
 	\subfigure[]{
	\includegraphics[width=0.49\textwidth]{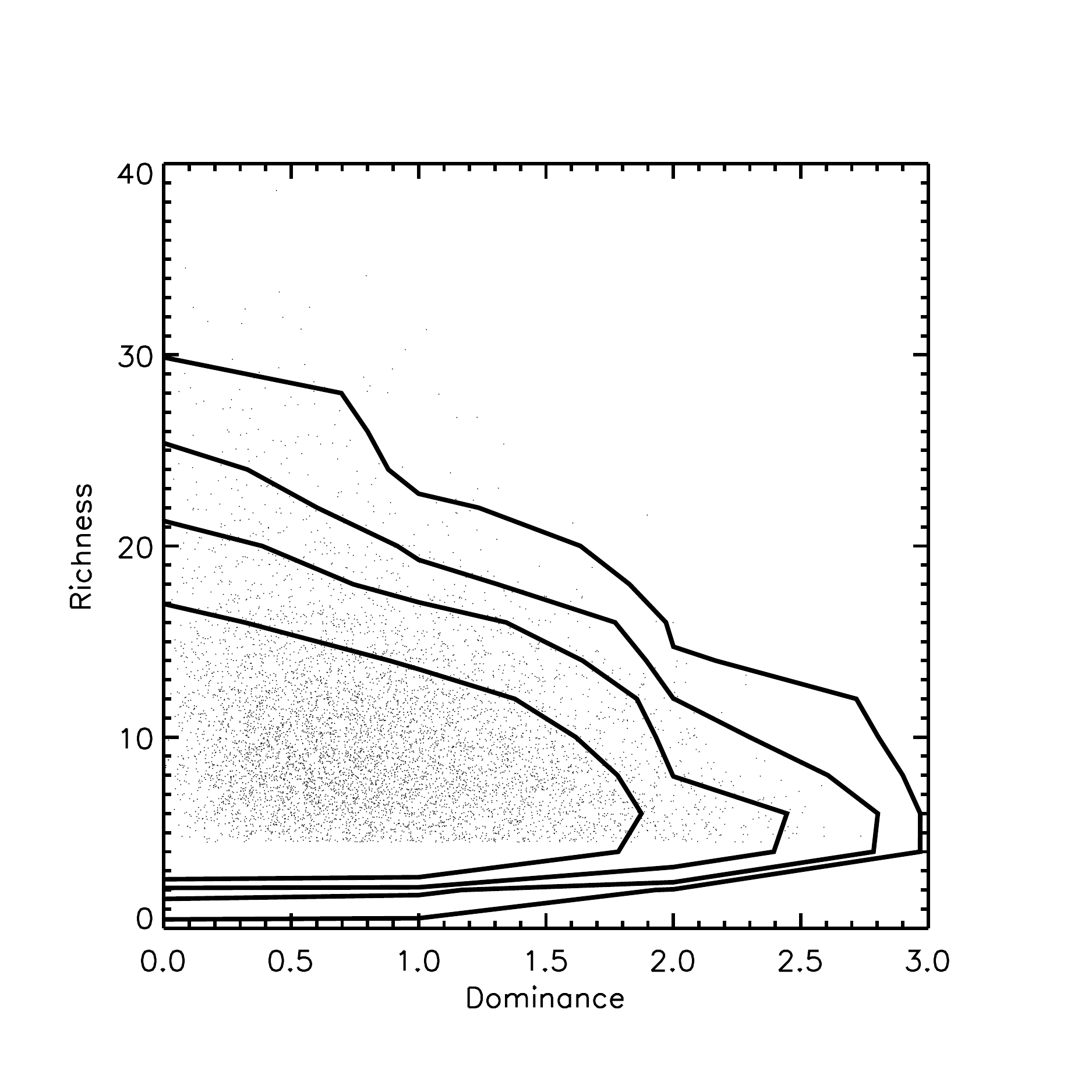}}
 	\subfigure[]{
	\includegraphics[width=0.49\textwidth]{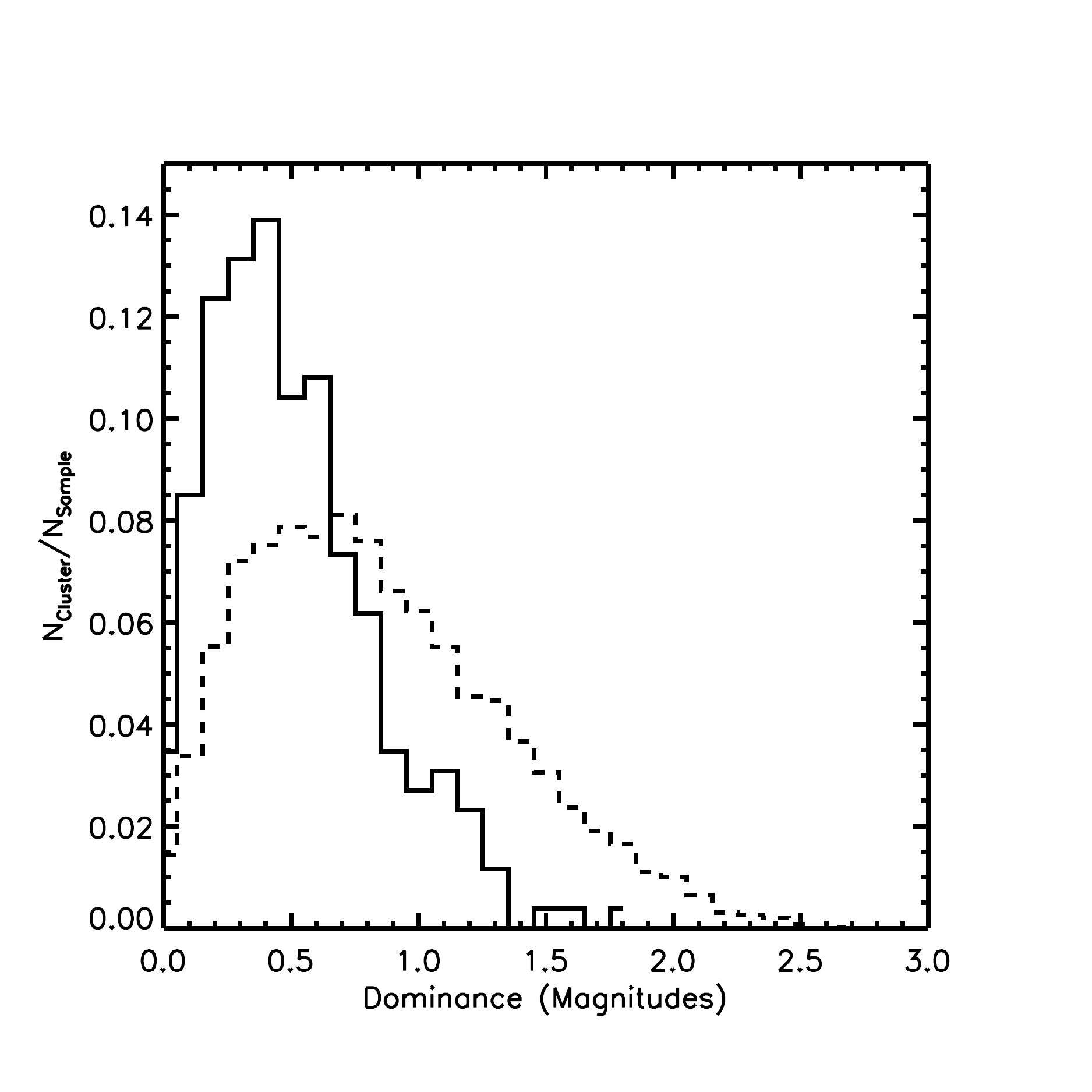}}
	\caption{ Panel (a) Scatter plot of cluster richness and BCG dominance, indicating that poorer clusters have more dominant BCGs. Panel (b) Histogram of BCG dominance split on our definition of rich and poor clusters. The BCGs of rich (solid line) clusters tend to be less dominant than those of poor (dashed line) clusters. In a rich cluster it is more probable that there is an unusually bright second ranked cluster galaxy than in poor clusters, assuming that both cluster galaxy populations have the same luminosity function.}
	\label{fig:domrichhisto}
\end{figure*}

\indent The denser environments of rich clusters offer more possibilities for galaxy-galaxy interactions during cluster formation and cluster evolution,which may affect the observed alignment. Figure \ref{fig:historich}a is a scatter plot of richness i.e. number of red sequence galaxies in a given cluster with alignment. There is a slight indication that richer clusters show stronger alignment. Figure \ref{fig:historich}b shows the histogram of alignment for clusters with fewer than and more than $20$ members, picking out the high-count tail of the cluster richness distribution (Figure \ref{fig:richhisto}). Rich clusters show a stronger alignment, a difference significant at the $2.3\sigma$ level. The rich clusters in the Centre subsample are even more strongly aligned. With the offset cut, the results are not significant (Table \ref{tab:richphi}). K-S tests show that the rich and poor samples in the Total and Centre cuts are different with high confidence, but the difference is statistically insignificant in the Offset sample (see Table \ref{tab:ksrich}).
\\
\indent We investigate the possibility that position angle determinations may be noisier in poor clusters than in rich clusters due to small number statistics which may suppress the alignment signal. This could explain the weaker alignment seen in poorer clusters.  In Figure \ref{fig:uncert_ba_rich} we show the statistical uncertainty on the cluster position angle ($\sigma_{\phi}$) as a function of cluster richness (left-hand panel)  and cluster axis ratio (right-hand panel). The main source of uncertainty in our determined cluster position angles stems from a cluster's axis ratio. With this in mind we investigate the axis ratio distributions of our rich and poor sub-samples. The mean axis ratio in the rich sub-sample is higher (i.e. the clusters are rounder) $\langle b/a \rangle=0.53$ than that in the poor sub-sample $\langle b/a \rangle=0.38$. Since a higher axis ratio will degrade the alignment signal more by introducing larger shot noise errors in the cluster's position angle, we are confident that the stronger alignment observed in the rich clusters is real.
\\
\indent Even though rich clusters tend to have more strongly aligned BCGs than poor clusters, their BCGs are \em less \rm dominant than those of the poor clusters (Figure \ref{fig:domrichhisto}). Assuming no correlation between cluster richness and the galaxy luminosity function \citep[see e.g..][]{Tremaine:1977p403,Loh:2006p425} it is more likely that a rich cluster has an uncommonly bright second ranked cluster galaxy, resulting in less dominant BCGs than in poor clusters.

\begin{table*}
\centering
\caption{Alignment between the BCG and cluster for rich ($N_{Member}\geq20$) and poor ($N_{Member}<20$) clusters separately.}
\begin{tabular}{@{}llcclcc@{}}
\hline
\null & \null& Dong et al.& \null \\
Sample & $N_{Cluster}$ & $\mathcal{R}$ & $\pm$ \\
\hline
Total Rich & 259& 1.023&0.120  \\
Total Poor&6772 &0.744 &0.018  \\
Centre Rich &141 &1.350 & 0.229  \\
Centre Poor & 3272& 0.849&0.030  \\
Offset Rich& 118&0.735 & 0.137 \\
Offset Poor& 3500& 0.656&0.023 \\
\hline
\hline
\null & \null & Dong et al. & \null & \null & \null  \\
\null & \null & ($0.1<z<0.3$) & \null & \null & Koester et al.  \\
Sample & $N_{Cluster}$ & $\mathcal{R}$ & $\pm$ & $N_{Cluster}$ & $\mathcal{R}$ & $\pm$  \\
\hline
Total Rich & 81&1.190 & 0.265 & 211& 1.110&0.153 \\
Total Poor& 2638& 0.823&0.032& 5533 & 0.849 & 0.023 \\
Centre Rich & 41& 1.563& 0.500& 135& 1.368&0.238  \\
Centre Poor & 1324& 0.961&0.053 &3512&0.992&0.033 \\
Offset Rich& 40&0.905& 0.286&76 &0.767 &0.178\\
Offset Poor& 1314& 0.704& 0.039&2021&0.644&0.029 \\
\end{tabular}
\label{tab:richphi}
\end{table*}

\begin{table}
\centering
\caption{Kolmogorov-Smirnov tests determining whether the rich and poor sub-samples are drawn from different underlying distributions. $\Delta$ gives the significance of the difference between $\mathcal{R}$ statistics for the rich and poor samples in units of standard deviation ($\sigma$).}
\begin{tabular}{@{}llcc@{}}
\hline
Sample & K-S Confidence&$\Delta$ \\
\hline
Dong Total \\ ($0.08<z<0.44$)  & $99.99\%$&$2.3$\\
Centre &$99.88\%$&$2.2$\\
Offset &$92.04\%$&$0.6$\\
\hline
Dong Total \\ ($0.1<z<0.3$) & $97.94\%$&$1.4$\\
Centre & $90.89\%$&$1.2$\\
Offset & $77.37\%$&$0.7$\\
\hline
Koester
Total & $96.53\%$&$1.7$\\
Centre & $96.88\%$&$1.6$\\
Offset & $43.98\%$&$0.7$ \\
\end{tabular}
\label{tab:ksrich}
\end{table}

\subsection{Redshift Evolution}
\begin{figure}
	\centering
	\includegraphics[width=0.5\textwidth]{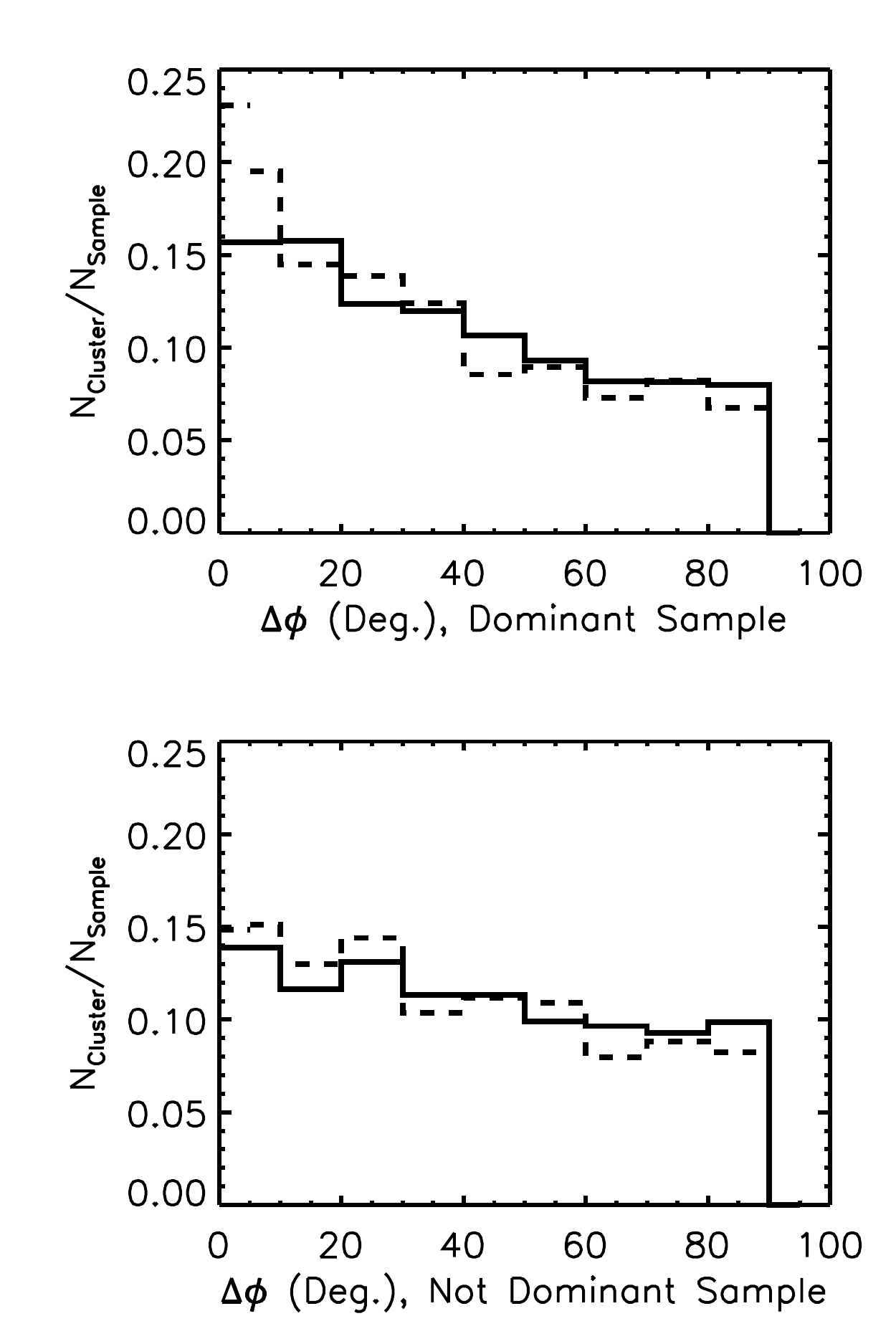}
	\caption{Histogram of cluster-BCG alignment examining the effect of BCG dominance and redshift on the alignment signal. The top panel shows the dominant sample and the bottom panel shows the non-dominant sample. The solid histograms show the high redshift ($z\geq0.26$) sample and the dashed histograms show the low redshift sample. The alignment increases by approximately $2.4\sigma$ from high to low z in the dominant case. In the non-dominant case there is no significant change in alignment.}
	\label{fig:histodomz}
\end{figure}

It is certainly possible that the mechanisms responsible for the alignment between BCGs and the cluster act over the full span of the cluster's lifetime rather than during or immediately after cluster virialization. For example, tidal torques exerted by the cluster or secondary infall may enhance or reduce any primordial alignment \citep[e.g..][]{Ciotti:1994p478,Hopkins:2005p348,Altay:2006p479}. It has been argued by \citet{Merritt:1985p78} and \citet{Tremaine:1990dig..book..394T} among others, that the observed dominance of BCGs cannot be achieved via cannibalism of other cluster members during the BCG's lifetime, since the high velocity dispersion of cluster galaxies makes frequent merging unlikely. Thus the BCG's dominance must already be in place shortly after the cluster has formed. Indeed, we observe no evolution in the distribution of BCG dominance from high to low redshift (at least for $0.08<z<0.44$). Thus if we observe any evolution of alignment during the cluster's lifetime it cannot be the result of the same mechanism that makes the BCGs dominant. To test this, we split our dominant and non-dominant samples into a high redshift and a low redshift sample, taking $z=0.26$ roughly the median of the Dong et al. sample as the dividing line. The median look-back times of the two sub-samples are $2.1$ Gyr and $3.7$ Gyr. We find that the mean axis ratios of the BCGs and the clusters are the same in the high and low redshift samples. Thus uncertainties  in measuring the BCG's shape stemming from lower signal-to-noise ratio photometry and a smaller physical size on the sky at high redshift, do not bias our alignment signal. 
\\
\indent Figure \ref{fig:histodomz} shows the alignment histograms for the Total dominant and non-dominant samples at high and low redshift. The alignment effect in the Total sample increases in $\mathcal{R}$ by approximately $3.1\sigma$ from high to low redshift. In the dominant sample we see an increase of roughly $2.4\sigma$; there is no statistically significant evolution in the non-dominant sample. Even though the $\mathcal{R}$ values for the high and low redshift samples are only $~2\sigma$ apart, K-S tests show at the $99\%$ confidence level that the high and low redshift samples are drawn from different underlying distributions in both the dominant and non-dominant samples for the Centre sample. The $\mathcal{R}$ statistic probes a specific aspect of the alignment distribution where as the K-S tests asks the more general question of similarity between distributions. Two different distributions can have the same $\mathcal{R}$ value but a K-S test will show that they are different. However, due to Poisson noise two similar distributions may have different $\mathcal{R}$ values yet the K-S test will show that they are in fact the same. Thus in this case it seems that even though the $\mathcal{R}$ values are only $~2\sigma$ apart the results of the K-S test are a stronger constraint on the redshift evolution. The statistics are summarised in Tables \ref{tab:lambdasumz} \ref{tab:lambdasumdomz}, and \ref{tab:ksredsh}.

\begin{table*}
\centering
\caption{Alignment between the BCG and cluster distinguishing between high redshift ($z\geq0.26$) and low redshift ($z<0.26$) clusters.}
\begin{tabular}{@{}llcclcc@{}}
\hline
\null & \null& $z\geq0.26$& \null & \null & $z<0.26$&\null  \\
Sample & $N_{Cluster}$ & $\mathcal{R}$ & $\pm$& $N_{Cluster}$ & $\mathcal{R}$ & $\pm$  \\
\hline
Total & 5036& 0.718&0.021&1991 &0.851&0.038 \\
Centre &2418&0.803&0.033 &992&1.037&0.066\\
Offset &2618&0.646&0.026&999&0.696&0.045 \\
\end{tabular}
\label{tab:lambdasumz}
\end{table*}

\begin{table*}
\centering
\caption{Alignment between the BCG and cluster distinguishing between dominant and non-dominant BCGs at high and low redshift.}
\begin{tabular}{@{}llcclcc@{}}
\hline
\null & \null& $z\geq0.26$& \null & \null & $z<0.26$&\null  \\
Sample & $N_{Cluster}$ & $\mathcal{R}$ & $\pm$& $N_{Cluster}$ & $\mathcal{R}$ & $\pm$  \\
\hline
Total Dominant&3077&0.779&0.028 &1276&0.919&0.051\\
Total Not Dominant &1959&0.630&0.029 &715&0.740&0.056\\
Centre Dominant &1567&0.890&0.045&689&1.082&0.082 \\
Centre Not Dominant&851&0.662&0.046 &303&0.942&0.108  \\
Offset Dominant &1510&0.676&0.035&587&0.757&0.063 \\
Offset Not Dominant &1108&0.606&0.038&412&0.616&0.062 \\
\end{tabular}
\label{tab:lambdasumdomz}
\end{table*}

\indent If we now split the Total sample on richness instead of dominance (Tables  \ref{tab:lambdasumrichz} and \ref{tab:ksredsh}), we find no statistically significant evolution in the alignment of the rich sub-sample. The Poisson errors are substantial, however, due to the small number of rich clusters in the sample. However, there is a trend toward stronger alignment at low redshift in the poor sample, which is significant at $3.2\sigma$.

\begin{table*}
\centering
\caption{Alignment between the BCG and cluster distinguishing between rich ($N_{Members}\geq20$) and poor ($N_{Member}<20$) clusters at high and low redshift.}
\begin{tabular}{@{}llcclcc@{}}
\hline
\null & \null& $z\geq0.26$& \null & \null & $z<0.26$&\null  \\
Sample & $N_{Cluster}$ & $\mathcal{R}$ & $\pm$& $N_{Cluster}$ & $\mathcal{R}$ & $\pm$  \\
\hline
Total Rich&203&1.051&0.148&56&0.931&0.249  \\
Total Poor&4836&0.706&0.021&1936&0.847&0.039 \\
Centre Rich&112&1.383&0.265&29&1.231&0.460 \\
Centre Poor&2308&0.782&0.033&964&1.029&0.066 \\
Offset Rich&91&0.750&0.159&27&0.688&0.269 \\
Offset Poor&2528&0.642&0.026 &972&0.696&0.045 \\
\end{tabular}
\label{tab:lambdasumrichz}
\end{table*}

\begin{table}
\centering
\caption{Kolmogorov-Smirnov tests determining whether the high-redshift and low-redshift sub-samples are drawn from different underlying distributions. $\Delta$ gives the significance of the difference between $\mathcal{R}$ statistics for the high and low redshift samples in units of standard deviation ($\sigma$).}
\begin{tabular}{@{}llcc@{}}
\hline
Sample & K-S Confidence & $\Delta$\\
\hline
Total & $99.14\%$& 3.1\\
Centre &$99.82\%$&3.2\\
Offset &$38.25\%$ &1.0\\
\hline
Total Dominant& $97.75\%$&2.4\\
Total Not Dominant&$80.45\%$&1.7 \\
\hline
Total Rich& $11.91\%$&0.4\\
Total Poor& $99.48\%$&3.2\\
\hline
Centre Dominant& $99.07\%$&2.1\\
Centre Not Dominant& $98.97\%$&2.4 \\
\hline
Centre Rich& $36.80\%$&0.3\\
Centre Poor& $99.83\%$&3.3\\
\hline
Offset Dominant& $46.37\%$&1.1\\
Offset Not Dominant& $49.27\%$&0.1\\
\hline
Offset Rich& $36.08\%$&0.2\\
Offset Poor& $38.66\%$&1.0
\end{tabular}
\label{tab:ksredsh}
\end{table}

\section{Summary and Discussion}
\indent Using the SDSS Data Release $6$ and the cluster catalogues developed by \citet{Dong:2008p220} ($7031$ clusters) and \citet{Koester:2007p485} ($5744$ clusters) we investigate the alignment between the BCG and cluster position angles, searching for influences from BCG dominance in luminosity, cluster richness, and redshift. Understanding the alignment offers an observational test of the current cosmological paradigm and also allows for a better understanding of contamination in lensing studies \citep{2006MNRAS.367..611M}. We have confirmed that BCGs are preferentially aligned with their host cluster and find that dominant BCGs have a $4.4\sigma$ stronger alignment signal than non-dominant ones. This suggests that dominance and alignment result from the same physical process. The sample of clusters in which our selected BCG is farther than $0.2$ Mpc from the cluster centre shows a $5.5\sigma$ weaker alignment signal, which may be an indication that these Offset clusters are the result of cluster mergers. If dynamical interactions influence alignment, one might expect a dependence of alignment on cluster richness. We found that rich clusters (defined as containing more than $20$ red sequence galaxies within $0.5$ Mpc from the cluster centre) show a stronger alignment signal than do poor clusters. The difference in alignment is only significant at the $2.3\sigma$ level, but this is due to the large Poisson errors given the small sample of rich clusters. A Kolmogorov-Smirnov test shows that the alignment distributions in the rich and poor clusters are different at better than $99\%$ confidence.
\\
\indent Our cluster sample extends out to $z=0.44$, allowing us to test redshift evolution. Non-linear effects during cluster virialization can weaken primordial alignments. Thus if the BCG has resided in the cluster for many crossing times its initial infall direction may have been forgotten. However, we find that the alignment signal is stronger at lower redshift, an effect significant at more than $3\sigma$ for those clusters whose BCGs lie within $0.2$ Mpc of the Centre. For the Offset clusters we do not find a statistically significant evolution in alignment over the redshift range considered here. 
\\
\indent A likely cause for the correlation between BCG dominance and its alignment with the host cluster galaxy distribution is that both are the result of galaxies falling along primordial filaments and merging during cluster formation \citep[see e.g.][]{West:1995p475,Fuller:1999p61,Torlina:2007p58, Garijo:1997p54, Dubinski:1998p366,Knebe:2004ApJ...603....7K}. Recent modelling using the Millennium Simulation \citep{Delucia:2007MNRAS.375....2D} suggests that the stars that eventually end up in the BCGs are formed early ($z\approx 4$) but that the final BCG is assembled comparatively, late with half their mass being locked up in a single galaxy only after $z\approx 0.5$. The stellar mass of BCGs grows by a factor of $3-4$ via mergers since $z=1$ \citep{Gao:2004p1302,AragonSalamanca:1998p1268,Delucia:2007MNRAS.375....2D}. Observational studies, however, suggest that $~80\%$ of the stars that make up massive ($>4L_{\star}$) red galaxies today are bound up in a single galaxy by $z=0.7$ and that the stellar mass in red galaxies evolves only by a factor of $~2$ since $z=1$. In the case of massive red galaxies this is mostly due to dry mergers \citep{Faber:2007p1433, Brown:2007ApJ...654..858B}. The observation that BCG dominance evolves little with redshift (at least in the range considered here) suggests that the mechanism that causes the BCG to become dominant acts early on in the cluster's history. However, as pointed out by \citet{Faltenbacher:2007p60} any primordial alignment on small scales is likely reduced by non-linear processes during cluster virialization. A possibility is that the BCG is somehow resistant to realignment after it has settled at the bottom of the potential well. Secondary infall episodes may reinforce the primordial alignments, which may be the cause of the slight redshift evolution of the alignment that we observe. If galaxy mergers early in a cluster's history are the cause of both BCG dominance (with more mergers creating more dominant BCGs) and its alignment with the host cluster, the uniformity of the BCG population as a whole remains puzzling, since BCGs in poorer clusters will have undergone fewer mergers than those in rich clusters. Perhaps the colour and luminosity of BCGs are not affected by mergers after a certain threshold of merging activity. Additional mergers then only affect the BCG orientation within the cluster removing other galaxies from the cluster's population (preferentially brighter more massive galaxies) letting the BCG grow in dominance.  Further investigations of clusters in the redshift range $0.5<z<1$ will be very useful in understanding how the growth in dominance of the BCG and its alignment are related.
\\ 
\indent Filamentary infall is likely to play an important role in determining BCG and cluster orientations early in cluster's history, but tidal torques from the large scale cluster mass distribution can influence the alignment signals between BCGs and clusters during the entire cluster lifetime. Simulations show that tides align a galaxy's major axis with the cluster's radial direction on timescales greater than the galaxy's dynamical time scale but significantly shorter than a Hubble time. The effect is strongest for galaxies outside the cluster core radius \citep{Ciotti:1994p478}, and is therefore less important for central BCGs. The simulations by \citet{Faltenbacher:2008p67} find on galaxy scales tides cause a radial alignment. Even though tidal torques cause some kinds of alignment they are not likely the source of the BCG-cluster alignment we observe. The evolution of the alignment signal needs to be better understood theoretically to differentiate the effects of tidal torques and infall episodes. These questions can likely be addressed using the Millennium Simulation which has already been used to explore properties of BCGs \citep[see e.g.][]{2009ApJ...696.1094R}
\\
\indent Recent surveys such as the SDSS Stripe 82 and future surveys such as LSST \citep[][LSST Science Book]{2009arXiv0912.0201L}  and PanSTARRS \citep{2002SPIE.4836..154K} with deeper photometry will allow us to probe clusters at higher redshifts to study the evolution of alignment over a larger range of cosmic time.  Deeper surveys will also increase the number of galaxies in low-redshift clusters, reducing Poisson noise in our samples. This will allow us to study cluster shapes and alignments for galaxies of different luminosities, a diagnostic of different accretion and dynamical histories. Simulations addressing the connection between BCG dominance and alignment will be important in discovering which mechanisms cause the distinct physical properties of the BCGs and their relation to their parent cluster.
\\
\\
MNO is funded by the Gates Cambridge Trust, the Isaac Newton Studentship fund and the Science and Technology Facilities Council (STFC). MAS was supported in part by NSF grant AST-0707266. We thank James E. Gunn and Michael D. Gladders for valuable discussions and feedback and the anonymous referee for help in clarifying the text.

Funding for the SDSS and SDSS-II has been provided by the Alfred P. Sloan Foundation, the Participating Institutions, the National Science Foundation, the U.S. Department of Energy, the National Aeronautics and Space Administration, the Japanese Monbukagakusho, the Max Planck Society, and the Higher Education Funding Council for England. The SDSS Web Site is http://www.sdss.org/.

\bibliography{citations_bcg}

\begin{thebibliography}{82}
\expandafter\ifx\csname natexlab\endcsname\relax\def\natexlab#1{#1}\fi

\bibitem[Adelman-McCarthy et~al.(2008)Adelman-McCarthy, Ag{\"u}eros, Allam
  et~al.]{AdelmanMcCarthy:2008p362}
Adelman-McCarthy J.~K., Ag{\"u}eros M.~A., Allam S.~S., et~al., 2008, ApJS,
  175, 297

\bibitem[{Allgood} et~al.(2006){Allgood}, {Flores}, {Primack}
  et~al.]{Allgood2006MNRAS.367.1781A}
{Allgood} B., {Flores} R.~A., {Primack} J.~R., et~al., 2006, \mnras, 367, 1781

\bibitem[Altay et~al.(2006)Altay, Colberg \& Croft]{Altay:2006p479}
Altay G., Colberg J.~M., Croft R. A.~C., 2006, \mnras, 370, 3, 1422

\bibitem[{Annis} et~al.(1999){Annis}, {Kent}, {Castander}
  et~al.]{Annis:1999AAS...195.1202A}
{Annis} J., {Kent} S., {Castander} F., et~al., 1999, in { Bulletin of the
  American Astronomical Society\/}, vol.~31,  1391

\bibitem[Aragon-Salamanca et~al.(1998)Aragon-Salamanca, Baugh \&
  Kauffmann]{AragonSalamanca:1998p1268}
Aragon-Salamanca A., Baugh C.~M., Kauffmann G., 1998, \mnras, 297, 427

\bibitem[Baldry et~al.(2004)Baldry, Glazebrook, Brinkmann
  et~al.]{Baldry:2004p481}
Baldry I.~K., Glazebrook K., Brinkmann J., et~al., 2004, \apj, 600, 681

\bibitem[Basilakos et~al.(2006)Basilakos, Plionis, Yepes, Gottlober \&
  Turchaninov]{Basilakos:2006p69}
Basilakos S., Plionis M., Yepes G., Gottlober S., Turchaninov V., 2006, \mnras,
  365, 2, 539

\bibitem[{Baum}(1959)]{Baum:1959PASP...71..106B}
{Baum} W.~A., 1959, \pasp, 71, 106

\bibitem[Bell et~al.(2004)Bell, Wolf, Meisenheimer et~al.]{Bell:2004p497}
Bell E.~F., Wolf C., Meisenheimer K., et~al., 2004, \apj, 608, 752

\bibitem[Binggeli(1982)]{Binggeli:1982p376}
Binggeli B., 1982, A\&A, 107, 338

\bibitem[Bramel et~al.(2000)Bramel, Nichol \& Pope]{Bramel:2000p520}
Bramel D.~A., Nichol R.~C., Pope A.~C., 2000, \apj, 533, 601

\bibitem[{Brown} et~al.(2007){Brown}, {Dey}, {Jannuzi}
  et~al.]{Brown:2007ApJ...654..858B}
{Brown} M.~J.~I., {Dey} A., {Jannuzi} B.~T., et~al., 2007, \apj, 654, 858

\bibitem[Carter \& Metcalfe(1980)]{Carter:1980p229}
Carter D., Metcalfe N., 1980, \mnras, 191, 325

\bibitem[Cassata et~al.(2008)Cassata, Cimatti, Kurk et~al.]{Cassata:2008p507}
Cassata P., Cimatti A., Kurk J., et~al., 2008, A\&A, 483, L39

\bibitem[Ciotti \& Dutta(1994)]{Ciotti:1994p478}
Ciotti L., Dutta S.~N., 1994, \mnras, 270, 390

\bibitem[{De Lucia} \& {Blaizot}(2007)]{Delucia:2007MNRAS.375....2D}
{De Lucia} G., {Blaizot} J., 2007, \mnras, 375, 2

\bibitem[{Djorgovski}(1987)]{Djorgovski1987}
{Djorgovski} S.~G., 1987, in { Nearly Normal Galaxies. From the Planck Time to
  the Present\/}, edited by {S.~M.~Faber},  227--233

\bibitem[Dong et~al.(2008)Dong, Pierpaoli, Gunn \& Wechsler]{Dong:2008p220}
Dong F., Pierpaoli E., Gunn J.~E., Wechsler R.~H., 2008, \apj, 676, 868

\bibitem[Donoso et~al.(2006)Donoso, O'Mill \& Lambas]{Donoso:2006p66}
Donoso E., O'Mill A., Lambas D.~G., 2006, \mnras, 369, 1, 479

\bibitem[Dubinski(1998)]{Dubinski:1998p366}
Dubinski J., 1998, \apj, 502, 141

\bibitem[Eisenhardt et~al.(2008)Eisenhardt, Brodwin, Gonzalez
  et~al.]{Eisenhardt:2008p1444}
Eisenhardt P. R.~M., Brodwin M., Gonzalez A.~H., et~al., 2008, \apj, 684, 905

\bibitem[Faber et~al.(2007)Faber, Willmer, Wolf et~al.]{Faber:2007p1433}
Faber S.~M., Willmer C. N.~A., Wolf C., et~al., 2007, \apj, 665, 265

\bibitem[Faltenbacher et~al.(2008)Faltenbacher, Jing, Li
  et~al.]{Faltenbacher:2008p67}
Faltenbacher A., Jing Y.~P., Li C., et~al., 2008, \apj, 675, 146

\bibitem[Faltenbacher et~al.(2007)Faltenbacher, Li, Mao
  et~al.]{Faltenbacher:2007p60}
Faltenbacher A., Li C., Mao S., et~al., 2007, \apj, 662, L71

\bibitem[{Faltenbacher} et~al.(2009){Faltenbacher}, {Li}, {White}, {Jing},
  {Shu-DeMao} \& {Wang}]{Faltenbacher2009RAA.....9...41F}
{Faltenbacher} A., {Li} C., {White} S.~D.~M., {Jing} Y., {Shu-DeMao}, {Wang}
  J., 2009, Research in Astronomy and Astrophysics, 9, 41

\bibitem[Fuller et~al.(1999)Fuller, West \& Bridges]{Fuller:1999p61}
Fuller T.~M., West M.~J., Bridges T.~J., 1999, \apj, 519, 22

\bibitem[Gao et~al.(2004)Gao, Loeb, Peebles, White \& Jenkins]{Gao:2004p1302}
Gao L., Loeb A., Peebles P. J.~E., White S. D.~M., Jenkins A., 2004, \apj, 614,
  17

\bibitem[Garijo et~al.(1997)Garijo, Athanassoula \&
  Garcia-Gomez]{Garijo:1997p54}
Garijo A., Athanassoula E., Garcia-Gomez C., 1997, A\&A, 327, 930

\bibitem[Gladders \& Yee(2000)]{Gladders:2000p360}
Gladders M.~D., Yee H. K.~C., 2000, \aj, 120, 2148

\bibitem[{Gladders} \& {Yee}(2005)]{Gladders:2005ApJS..157....1G}
{Gladders} M.~D., {Yee} H.~K.~C., 2005, \apjs, 157, 1

\bibitem[Gunn et~al.(1998)Gunn, Carr, Rockosi et~al.]{Gunn:1998p1435}
Gunn J.~E., Carr M., Rockosi C., et~al., 1998, \aj, 116, 3040

\bibitem[{Gunn} et~al.(2006){Gunn}, {Siegmund}, {Mannery}
  et~al.]{2006AJ....131.2332G}
{Gunn} J.~E., {Siegmund} W.~A., {Mannery} E.~J., et~al., 2006, \aj, 131, 2332

\bibitem[{Hashimoto} et~al.(2008){Hashimoto}, {Henry} \&
  {Boehringer}]{Hashimoto:2008MNRAS.390.1562H}
{Hashimoto} Y., {Henry} J.~P., {Boehringer} H., 2008, \mnras, 390, 1562

\bibitem[Hopkins et~al.(2005)Hopkins, Bahcall \& Bode]{Hopkins:2005p348}
Hopkins P.~F., Bahcall N.~A., Bode P., 2005, \apj, 618, 1

\bibitem[Humason et~al.(1956)Humason, Mayall \& Sandage]{Humason:1956p416}
Humason M.~L., Mayall N.~U., Sandage A.~R., 1956, \aj, 61, 97

\bibitem[Ivezi{\'c} et~al.(2004)Ivezi{\'c}, Lupton, Schlegel
  et~al.]{Ivezic:2004p1443}
Ivezi{\'c} {\v Z}., Lupton R.~H., Schlegel D., et~al., 2004, Astronomische
  Nachrichten, 325, 583

\bibitem[{Kaiser} et~al.(2002){Kaiser}, {Aussel}, {Burke}
  et~al.]{2002SPIE.4836..154K}
{Kaiser} N., {Aussel} H., {Burke} B.~E., et~al., 2002, in { Society of
  Photo-Optical Instrumentation Engineers (SPIE) Conference Series\/}, edited
  by {J.~A.~Tyson \& S.~Wolff}, vol. 4836 of { Presented at the Society of
  Photo-Optical Instrumentation Engineers (SPIE) Conference\/},  154--164

\bibitem[{Kawasaki} et~al.(1998){Kawasaki}, {Shimasaku}, {Doi} \&
  {Okamura}]{Kawasaki:1998A&AS..130..567K}
{Kawasaki} W., {Shimasaku} K., {Doi} M., {Okamura} S., 1998, \aaps, 130, 567

\bibitem[{Kepner} et~al.(1999){Kepner}, {Fan}, {Bahcall}, {Gunn}, {Lupton} \&
  {Xu}]{Kepner:1999ApJ...517...78K}
{Kepner} J., {Fan} X., {Bahcall} N., {Gunn} J., {Lupton} R., {Xu} G., 1999,
  \apj, 517, 78

\bibitem[{Kim} et~al.(2002){Kim}, {Annis}, {Strauss} \&
  {Lupton}]{Kim:2002ASPC..268..395K}
{Kim} R.~S.~J., {Annis} J., {Strauss} M.~A., {Lupton} R.~H., 2002, in { Tracing
  Cosmic Evolution with Galaxy Clusters\/}, edited by {S.~Borgani, M.~Mezzetti,
  \& R.~Valdarnini}, vol. 268 of { Astronomical Society of the Pacific
  Conference Series\/},  395

\bibitem[Kim et~al.(2002)Kim, Kepner, Postman et~al.]{Kim:2002p370}
Kim R. S.~J., Kepner J.~V., Postman M., et~al., 2002, \aj, 123, 20

\bibitem[{Knebe} et~al.(2004){Knebe}, {Gill}, {Gibson}, {Lewis}, {Ibata} \&
  {Dopita}]{Knebe:2004ApJ...603....7K}
{Knebe} A., {Gill} S.~P.~D., {Gibson} B.~K., {Lewis} G.~F., {Ibata} R.~A.,
  {Dopita} M.~A., 2004, \apj, 603, 7

\bibitem[Koester et~al.(2007{\natexlab{a}})Koester, McKay, Annis
  et~al.]{Koester:2007p484}
Koester B.~P., McKay T.~A., Annis J., et~al., 2007{\natexlab{a}}, \apj, 660,
  221

\bibitem[Koester et~al.(2007{\natexlab{b}})Koester, McKay, Annis
  et~al.]{Koester:2007p485}
Koester B.~P., McKay T.~A., Annis J., et~al., 2007{\natexlab{b}}, \apj, 660,
  239

\bibitem[Komatsu et~al.(2009)Komatsu, Dunkley, Nolta et~al.]{Komatsu:2009p1215}
Komatsu E., Dunkley J., Nolta M.~R., et~al., 2009, \apjs, 180, 2, 330

\bibitem[Lambas et~al.(1988)Lambas, Groth \& Peebles]{Lambas:1988p43}
Lambas D.~G., Groth E.~J., Peebles P. J.~E., 1988, \aj, 95, 996

\bibitem[Lauer(1988)]{Lauer:1988p402}
Lauer T.~R., 1988, \apj, 325, 49

\bibitem[Loh \& Strauss(2006)]{Loh:2006p425}
Loh Y.-S., Strauss M.~A., 2006, \mnras, 366, 373

\bibitem[{LSST Science Collaborations: Paul A.~Abell} et~al.(2009){LSST Science
  Collaborations: Paul A.~Abell}, {Allison}, {Anderson}
  et~al.]{2009arXiv0912.0201L}
{LSST Science Collaborations: Paul A.~Abell}, {Allison} J., {Anderson} S.~F.,
  et~al., 2009, ArXiv e-prints

\bibitem[{Lubin} et~al.(2000){Lubin}, {Brunner}, {Metzger}, {Postman} \&
  {Oke}]{Lubin:2000ApJ...531L...5L}
{Lubin} L.~M., {Brunner} R., {Metzger} M.~R., {Postman} M., {Oke} J.~B., 2000,
  \apjl, 531, L5

\bibitem[Lupton et~al.(2001)Lupton, Gunn, Ivezi{\'c}, Knapp \&
  Kent]{Lupton:2001p1436}
Lupton R., Gunn J.~E., Ivezi{\'c} Z., Knapp G.~R., Kent S., 2001, Astronomical
  Data Analysis Software and Systems X, 238, 269

\bibitem[{Lupton} et~al.(2002){Lupton}, {Ivezic}, {Gunn}, {Knapp}, {Strauss} \&
  {Yasuda}]{Lupton:2002SPIE.4836..350L}
{Lupton} R.~H., {Ivezic} Z., {Gunn} J.~E., {Knapp} G., {Strauss} M.~A.,
  {Yasuda} N., 2002, in { Society of Photo-Optical Instrumentation Engineers
  (SPIE) Conference Series\/}, edited by {J.~A.~Tyson \& S.~Wolff}, vol. 4836,
  350--356

\bibitem[{Mandelbaum} et~al.(2006){Mandelbaum}, {Hirata}, {Ishak}, {Seljak} \&
  {Brinkmann}]{2006MNRAS.367..611M}
{Mandelbaum} R., {Hirata} C.~M., {Ishak} M., {Seljak} U., {Brinkmann} J., 2006,
  \mnras, 367, 611

\bibitem[Merritt(1985)]{Merritt:1985p78}
Merritt D., 1985, \apj, 289, 18

\bibitem[Navarro et~al.(1996)Navarro, Frenk \& White]{Navarro:1996p399}
Navarro J.~F., Frenk C.~S., White S. D.~M., 1996, \apj, 462, 563

\bibitem[Pier et~al.(2003)Pier, Munn, Hindsley et~al.]{Pier:2003p1437}
Pier J.~R., Munn J.~A., Hindsley R.~B., et~al., 2003, \aj, 125, 1559

\bibitem[Postman \& Lauer(1995)]{Postman:1995p237}
Postman M., Lauer T.~R., 1995, \apj, 440, 28

\bibitem[{Postman} et~al.(1996){Postman}, {Lubin}, {Gunn}
  et~al.]{Postman:1996AJ....111..615P}
{Postman} M., {Lubin} L.~M., {Gunn} J.~E., et~al., 1996, \aj, 111, 615

\bibitem[Rhee \& Katgert(1987)]{Rhee:1987p231}
Rhee G. F. R.~N., Katgert P., 1987, A\&A, 183, 217

\bibitem[{Ruszkowski} \& {Springel}(2009)]{2009ApJ...696.1094R}
{Ruszkowski} M., {Springel} V., 2009, \apj, 696, 1094

\bibitem[Sastry(1968)]{Sastry:1968p242}
Sastry G.~N., 1968, \pasp, 80, 252

\bibitem[Schlegel et~al.(1998)Schlegel, Finkbeiner \&
  Davis]{Schlegel:1998p1216}
Schlegel D.~J., Finkbeiner D.~P., Davis M., 1998, \apj, 500, 525

\bibitem[Schuecker \& Boehringer(1998)]{Schuecker:1998p521}
Schuecker P., Boehringer H., 1998, A\&A, 339, 315

\bibitem[Siverd et~al.(2009)Siverd, Ryden \& Gaudi]{Siverd:2009p516}
Siverd R.~J., Ryden B.~S., Gaudi B.~S., 2009, arXiv/astro-ph.GA-0903.2264v1

\bibitem[{Skibba} et~al.(2010){Skibba}, {van den Bosch}, {Yang}, {Mo}, {More}
  \& {Fontanot}]{Skibba2010AAS...21533002S}
{Skibba} R.~A., {van den Bosch} F., {Yang} X., {Mo} H., {More} S., {Fontanot}
  F., 2010, in { Bulletin of the American Astronomical Society\/}, vol.~41 of {
  Bulletin of the American Astronomical Society\/},  428--+

\bibitem[Splinter et~al.(1997)Splinter, Melott, Linn, Buck \&
  Tinker]{Splinter:1997p81}
Splinter R.~J., Melott A.~L., Linn A.~M., Buck C., Tinker J., 1997, \apj, 479,
  632

\bibitem[Stoughton et~al.(2002)Stoughton, Lupton, Bernardi, Blanton, Burles \&
  Castander]{Stoughton:2002p359}
Stoughton C., Lupton R.~H., Bernardi M., Blanton M.~R., Burles S., Castander
  F.~J., 2002, \aj, 123, 485

\bibitem[Strateva et~al.(2001)Strateva, Ivezi{\'c}, Knapp
  et~al.]{Strateva:2001p504}
Strateva I., Ivezi{\'c} {\v Z}., Knapp G.~R., et~al., 2001, \aj, 122, 1861

\bibitem[Struble(1990)]{Struble:1990p48}
Struble M.~F., 1990, \aj, 99, 743

\bibitem[Struble \& Peebles(1985)]{Struble:1985p234}
Struble M.~F., Peebles P. J.~E., 1985, \aj, 90, 582

\bibitem[Torlina et~al.(2007)Torlina, Propris \& West]{Torlina:2007p58}
Torlina L., Propris R.~D., West M.~J., 2007, \apj, 660, L97

\bibitem[{Tremaine}(1990)]{Tremaine:1990dig..book..394T}
{Tremaine} S., 1990, in Wielen, R., ed., Dynamics and Interactions of Galaxies,
   394--405, Springer Verlag, Berlin

\bibitem[Tremaine \& Richstone(1977)]{Tremaine:1977p403}
Tremaine S.~D., Richstone D.~O., 1977, \apj, 212, 311

\bibitem[Trevese et~al.(1992)Trevese, Cirimele \& Flin]{Trevese:1992p39}
Trevese D., Cirimele G., Flin P., 1992, \aj, 104, 935

\bibitem[Tucker et~al.(2006)Tucker, Kent, Richmond et~al.]{Tucker:2006p1438}
Tucker D.~L., Kent S., Richmond M.~W., et~al., 2006, Astronomische Nachrichten,
  327, 821

\bibitem[{Uomoto} et~al.(1999){Uomoto}, {Smee}, {Rockosi} et~al.]{Uomoto1999}
{Uomoto} A., {Smee} S., {Rockosi} C., et~al., 1999, in { Bulletin of the
  American Astronomical Society\/}, vol.~31 of { Bulletin of the American
  Astronomical Society\/},  1501--+

\bibitem[Wang et~al.(2008)Wang, Yang, Mo et~al.]{Wang:2008p65}
Wang Y., Yang X., Mo H.~J., et~al., 2008, \mnras, 385, 3, 1511

\bibitem[West et~al.(1995)West, Jones \& Forman]{West:1995p475}
West M.~J., Jones C., Forman W., 1995, \apj, 451, L5

\bibitem[Yang et~al.(2006)Yang, Bosch, Mo et~al.]{Yang:2006p82}
Yang X., Bosch F. C. V.~D., Mo H.~J., et~al., 2006, \mnras, 369, 1293

\bibitem[{Yasuda} et~al.(2001){Yasuda}, {Fukugita}, {Narayanan}
  et~al.]{2001AJ....122.1104Y}
{Yasuda} N., {Fukugita} M., {Narayanan} V.~K., et~al., 2001, \aj, 122, 1104

\bibitem[{Yee} et~al.(1999){Yee}, {Gladders} \&
  {L{\'o}pez-Cruz}]{Yee:1999ASPC..191..166Y}
{Yee} H.~K.~C., {Gladders} M.~D., {L{\'o}pez-Cruz} O., 1999, in { Photometric
  Redshifts and the Detection of High Redshift Galaxies\/}, edited by
  {R.~Weymann, L.~Storrie-Lombardi, M.~Sawicki, \& R.~Brunner}, vol. 191 of {
  Astronomical Society of the Pacific Conference Series\/},  166

\bibitem[York et~al.(2000)York, Adelman, Anderson et~al.]{York:2000p367}
York D.~G., Adelman J., Anderson J.~E., et~al., 2000, \aj, 120, 1579

\end{thebibliography}

\label{lastpage}

\end{document}